\documentclass[11pt,onecolumn]{IEEEtran}
\usepackage{amsthm}
\usepackage{times,amssymb,amsmath,amsfonts,float,nicefrac,color,bbm,mathrsfs,caption,float}
\usepackage{algorithm,enumerate,multirow,caption,tikz,graphicx}
\usepackage[mathscr]{eucal}
\usepackage{sidecap}
\usepackage{algpseudocode}
\usepackage{verbatim}
\usepackage{epstopdf}
\usepackage{textcomp}
\usetikzlibrary{shapes,arrows}
\usepackage{stmaryrd}
\usepackage{mathabx}
\usepackage[noadjust]{cite}
\usepackage{booktabs}
\usepackage[normalem]{ulem}
\usepackage[top=1in,bottom=1in,left=1in,right=1in]{geometry}
\allowdisplaybreaks

\newcommand{\RN}[1]{%
  \textup{\expandafter{\romannumeral#1}}%
}

\newcommand\remove[1]{}

\allowdisplaybreaks

\newtheorem{theorem}{Theorem}
\newtheorem{definition}{Definition}
\newtheorem{proposition}{Proposition}

\newtheorem{claim}{Claim}
\newtheorem{lemma}{Lemma}

\newtheorem{remark}{Remark}

\newcommand{\ff}{{\mathbb F}}

\def\mathbi#1{{\textbf{\textit #1}}}

\newcommand{\cC}{\mathcal{C}}

\newcommand{\cF}{\mathcal{F}}

\newcommand{\cR}{\mathcal{R}}

\DeclareMathOperator{\trace}{tr}

\DeclareMathOperator{\spun}{Span}

\begin{document}
\title{Repairing Reed-Solomon codes: Universally achieving the cut-set bound for any number of erasures}

\author{\IEEEauthorblockN{Min Ye} \hspace*{1in}
\and \IEEEauthorblockN{Alexander Barg}}

\maketitle
{\renewcommand{\thefootnote}{}\footnotetext{

\vspace{-.2in}
 
\noindent\rule{1.5in}{.4pt}

{M. Ye and A. Barg are with Dept. of ECE and ISR, University of Maryland, College Park, MD 20742. Emails: yeemmi@gmail.com and abarg@umd.edu. Their research is supported by NSF grants CCF1422955 and CCF1618603.}}
}
\renewcommand{\thefootnote}{\arabic{footnote}}
\setcounter{footnote}{0}

\begin{abstract}
The repair bandwidth of a code is the minimum amount of data required to repair one or several failed nodes (erasures). For MDS codes,
the repair bandwidth is bounded below by the so-called cut-set bound, and codes that meet this bound with equality are said
to support optimal repair of one or multiple failed nodes.

We consider the problem of repairing multiple failed nodes of Reed-Solomon (RS) codes.
In a recent work with I. Tamo (Proc. IEEE FOCS 2017), we gave the first explicit construction of RS codes with optimal repair of 
any single failed node from any subset of helper nodes.
In this paper, we construct explicit RS codes that universally achieve the cut-set bound for the repair of any number of failed nodes from any set of helper nodes. Moreover, the node size of our codes is close to the optimal (smallest possible) node size of codes with such property.
\end{abstract}

\section{Introduction}\label{section:repairing RS}

\subsection{Minimum Storage Regenerating codes and optimal repair bandwidth}
 The problem considered in this paper is motivated by the distributed nature of the system wherein the coded data is distributed across a large number of physical storage nodes. When some storage nodes fail,
the repair task performed by the system relies on communication between individual nodes, which introduces new challenges in the code design. In particular, a new parameter that has a bearing on the overall efficiency of the system is the {\em repair bandwidth}, i.e., the amount of data communicated between the nodes in the process of repairing failed nodes. 

Modern large-scale distributed storage systems rely on information encoding using Maximum Distance Separable (MDS) codes since they provide the optimal tradeoff between failure tolerance and storage overhead. 
To encode information with an MDS code, we represent data chunks as elements of a finite field. More specifically, we divide the original file into $k$ information blocks and view each block as a single element of a finite field $F$ or a vector over $F$. We encode the data by adding $r=n-k$ parity blocks (field symbols or vectors) and distribute the resulting $n$ blocks across $n$ storage nodes. The MDS property ensures that the original file can be recovered from the content stored on any $k$ nodes. In this paper we deal only with linear codes, so the parity blocks are formed as linear combinations of the information blocks over $F.$ We use the notation $(n,k)$ to refer to the length and dimension of a linear code.

Dimakis et. al. \cite{Dimakis10} gave a lower bound on the repair bandwidth of MDS codes for the repair of a single node failure, and Cadambe et. al. \cite{Cadambe13} generalized this bound to the repair of multiple node failures. Both these results
are now known as the {\em cut-set bound} on the repair bandwidth. MDS codes that achieve the cut-set bound with equality are called {\em minimum storage regenerating} (MSR) codes, and they
have been a focal point of current research in coding theory following their introduction in \cite{Dimakis10}.

Most studies of MDS codes with optimal repair bandwidth in the literature are concerned with a particular subclass of codes known as MDS {\em array codes} \cite{Blaum98}. Codewords of an $(n,k,l)$ 
MDS array code over a finite field $F$ have $k$ information nodes and $r=n-k$ parity nodes with the property that the contents
of any $k$ out of $n$ nodes suffices to recover the codeword.  Every
node is a column vector in $F^l,$ reflecting the fact that the system
views a large data block stored in one node as one coordinate of the codeword.
The parameter $l$ that determines the dimension
of each node is called {\em sub-packetization}.

Throughout the paper we use the notation $[n]:=\{1,2,\dots,n\}.$
Consider an $(n,k,l)$ array code $\cC$ over a finite field $F$. We write a codeword of $\cC$ as $c=(c_1,\dots,c_n)$,
where $c_i=(c_{i,0},c_{i,1},\dots,c_{i,l-1})^T\in F^l, i=1,\dots, n$. 
A node $i\in [n]$ can be repaired from a subset of $d\ge k$ helper nodes $\cR\subseteq[n]\backslash \{i\},$
by downloading $\beta_i(\cR)$ symbols of $F$ if there are
numbers $\beta_{ij}, j\in\cR$,
functions $f_{ij}: F^l\to F^{\beta_{ij}}, j\in\cR,$
and a function $g_i: F^{\sum_{j\in\cR}\beta_{ij}}\to F^l$
such that
   $$
   c_i=g_i(\{f_{ij}(c_{j}),j\in\cR \})
\text{~for all~} c=(c_1,\dots,c_n)\in \cC
   $$
   and
   $$
\sum_{j\in\cR}\beta_{ij}=\beta_i(\cR).
   $$
This definition extends straightforwardly to the repair of a subset of failed nodes $\cF\subseteq [n]$ from a subset of helper nodes $\cR\subseteq[n]\backslash \cF$.
%\begin{definition} [Repair scheme] \label{def:repair}
%Consider an $(n,k,l)$ array code $\cC$ over a finite field $F$. We write a codeword of $\cC$ as $c=(c_1,\dots,c_n)$,
%where $c_i=(c_{i,0},c_{i,1},\dots,c_{i,l-1})^T\in F^l, i=1,\dots, n$. 
%A node $i\in [n]$ can be repaired from a subset of $d\ge k$ helper nodes $\cR\subset[n]\backslash \{i\},$
%by downloading $\beta_i(\cR)$ symbols of $F$ if there are
%numbers $\beta_{ij}, j\in\cR$,
%functions $f_{ij}: F^l\to F^{\beta_{ij}}, j\in\cR,$
%and a function $g_i: F^{\sum_{j\in\cR}\beta_{ij}}\to F^l$
%such that
%   $$
%   c_i=g_i(\{f_{ij}(c_{j}),j\in\cR \})
%\text{~for all~} c=(c_1,\dots,c_n)\in \cC
%   $$
%   and
%   $$
%\sum_{j\in\cR}\beta_{ij}=\beta_i(\cR).
%   $$
%\end{definition}
We note that the symbols downloaded to repair the failed node(s) can be some functions of the contents  of the helper
nodes $c_j,j\in{\cR}$.

\begin{definition}[Repair bandwidth]
Let $\cC$ be an $(n,k,l)$ {MDS} array code over a finite field $F$ and let $c=(c_1,\dots,c_n)\in \cC$ be a codeword.
Given two disjoint subsets ${\cF},{\cR}\subseteq[n]$ such that $|{\cF}|\le r$ and $|{\cR}|\ge k,$ we define $N(\cC,{\cF},{\cR})$ as the smallest number of symbols of $F$ one needs to download from the helper nodes $\{c_i,i\in{\cR}\}$ in order to recover the failed (erased) 
nodes $\{c_i,i\in{\cF}\}.$  The \emph{$(h,d)$-repair bandwidth} of the code $\cC$ equals 
    \begin{equation}\label{eq:beta}
    \beta(h,d):=\max_{|{\cF}|=h,|{\cR}|=d, {\cF}\bigcap{\cR}=\emptyset} N(\cC,{\cF},{\cR}).
    \end{equation}
\end{definition}

The following basic result sets a benchmark for the minimum repair bandwidth.
\begin{theorem}[Cut-set bound \cite{Dimakis10,Cadambe13}] \label{def:csb}
Let $\cC$ be an $(n,k,l)$ MDS array code. For any two disjoint subsets ${\cF},{\cR}\subseteq[n]$ such that $|{\cF}|\le r$ and $|{\cR}|\ge k,$ we have the following inequality:
\begin{equation}\label{eq:cutset}
N(\cC,{\cF},{\cR})\ge \frac{|{\cF}||{\cR}|l}{|{\cF}|+|{\cR}|-k}.
\end{equation}
\end{theorem}

\begin{definition}
We say that an $(n,k,l)$ MDS code $\cC$ has the \emph{$(h,d)$-optimal repair} property if the $(h,d)$-repair bandwidth of $\cC$ (see \eqref{eq:beta})
equals 
   \begin{equation}\label{eq:cs1}
   \beta(h,d)=\frac{hdl}{h+d-k},
   \end{equation}
meeting the lower bound in \eqref{eq:cutset} with equality.
 \end{definition}
 
Another important parameter is the value of sub-packetization $l$. Due to the limited storage capacity of each node, we would like 
$l$ to be as small as possible. At the same time, $l$ cannot be too small; namely, as shown in \cite{Goparaju14}, for an $(n,k,d=n-1,l)$ MSR array code,  $l\ge 
2^{\sqrt{k/(2r-1)}}$.  
 
Several constructions of MDS array codes with optimal repair property are available in the literature. For the case of low 
code rate where $k\le n/2$, optimal-repair codes were constructed in \cite{Rashmi11}. For the high-rate regime see
\cite{Ye16,Goparaju17,Ye16a,Raviv17,Tamo13}. In particular, \cite{Ye16} gave explicit constructions of MDS array codes with the universal $(h,d)$-optimal repair property for all $h\le r$ and all $k\le d\le n-h$ simultaneously. In other words, the codes in \cite{Ye16} can repair any number of erasures $h$ from any set of $d$ helper nodes with the repair bandwidth achieving the cut-set bound \eqref{eq:cs1}.
 Recently the concept of repair bandwidth was extended in \cite{Tamo17} to the problem of correcting errors; \cite{Tamo17} also
presented explicit code constructions that support error correction under the minimum possible amount of information downloaded during the decoding process.

 \subsection{Repairing Reed-Solomon codes} \label{Sect:rty}
While there has been much research into constructions and properties of MDS array codes specifically designed for the repair task, it is
also of interest to study the repair bandwidth of well-known MDS codes, for instance, Reed-Solomon (RS) codes.
In \cite{Shanmugam14}, Shanmugam et al. proposed a framework for studying the repair bandwidth of a scalar linear $(n,k)$ MDS code $\cC$ 
over some finite field $E$ (called the symbol field below). The idea of \cite{Shanmugam14} is to ``vectorize'' the code construction by considering $\cC$ 
as an array code over some subfield $F$ of $E$. This approach provides a bridge between scalar MDS codes and MDS array codes, 
wherein the extension degree $l:=[E:F]$ can be viewed as the value of sub-packetization\footnote{The situation may in fact be
more involved: namely, the repair schemes for different nodes $i$ can be performed over different subfields $F_i$ of 
the field $F$. For instance, this is the case in \cite{Tamo17RS} for Reed-Solomon codes. 
In this case, it is a priori unclear what is the value of $l$, and we define it by isolating the largest subfield $L$ of $F$ with the 
property that the repair schemes for every node can be performed over it. In this case, the extension degree $l:=[F:L]$
is viewed as the subpacketization value of the code $\cC$.}. The code $\cC$ is viewed as an $(n,k,l)$ MDS array 
code over the field $F$, and the repair bandwidth is defined in exactly the same way as above. The cut-set bound 
\eqref{eq:cutset}-\eqref{eq:cs1} and the definition of the $(h,d)$-optimal repair property also apply to this setup.

In this paper we study the repair problem of RS codes, focusing on linear repair schemes, i.e., we assume that the repair operations
are linear over the field  $F.$
For the case of single node failure, Guruswami and Wootters \cite{Guruswami16} gave a characterization for
linear repair schemes of scalar linear MDS codes based on the framework in \cite{Shanmugam14}.
In \cite{Guruswami16}, the authors also gave explicit constructions of RS codes that can be repaired with smaller repair bandwidth than under the trivial approach.
Subsequently, the present authors \cite{Ye16b} used the general linear repair scheme in 
\cite{Guruswami16} to construct an explicit family of RS codes with asymptotically optimal repair bandwidth, and very recently
Chowdhury and Vardy \cite{Chowdhury17} further developed the results of \cite{Ye16b,Ye16}.
In \cite{Dau17}, Dau and Milenkovic generalized the scheme in \cite{Guruswami16} and extended their
results to a larger set of parameters.
Several works also extended the framework of \cite{Guruswami16} to repair more than one erasure (node failure) for RS codes \cite{Dau16,Bartan17}. At the same time, \cite{Guruswami16} as well as follow-up papers stopped short of constructing RS codes (or any scalar MDS codes) that meet the cut-set bound \eqref{eq:cs1} with equality (no matter for repairing single erasure or multiple erasures).
All the previous papers (apart from \cite{Ye16b}) focused on small sub-packetization regime, and the repair bandwidth of their constructions is rather far from the cut-set bound.

Very recently, Tamo and the present authors \cite{Tamo17RS} gave the first explicit construction of $(n,k)$ RS codes with $(1,d)$-optimal repair property for any given $k< d<n$.
The sub-packetization value of this construction is $l=\exp((1+o(1))n\log n)$. The authors of \cite{Tamo17RS} also proved an almost matching lower bound on $l$, showing that for scalar MDS codes (including the RS codes) to meet the cut-set bound with linear repair scheme, the sub-packetization $l$ must satisfy 
\begin{equation}\label{eq:lbl}
l \ge \exp((1+o(1)) k\log k).
\end{equation}

In this paper, we extend the construction in \cite{Tamo17RS} to the repair of multiple erasures. More precisely, given any $n>k$, we construct explicit $(n,k)$ RS codes with the universal $(h,d)$-optimal repair property for all $h\le r$ and all $k\le d \le n-h$ simultaneously. In other words, our codes can repair any number of failed nodes from any set of helper nodes with repair bandwidth achieving the cut-set bound.

The value of sub-packetization $l$ of our construction equals $r!$ times the product of the first $n$ 
distinct primes in an arithmetic progression, 
  \begin{equation}\label{eq:psi}
   l= r! \prod_{\substack{i=1\\[.02in]p_i\equiv 1\text{ mod\,}(r!)}}^n p_i.
  \end{equation}
As in \cite{Tamo17RS}, we invoke classic results of analytic number theory to describe the behavior of \eqref{eq:psi} for large $n$. In particular, the prime number theorem in arithmetic progressions (for instance, \cite[p.121]{IK04}) yields asymptotic estimates for $l$; see \cite{Tamo17RS} for a more detailed discussion. For fixed $r$ and growing $n$, we have $l= e^{(1+o(1)) n\log n}$, which is asymptotically the same as the result of \cite{Tamo17RS}.
According to the lower bound \eqref{eq:lbl}, when the code rate $k/n$ is close to $1$, the sub-packetization value of our codes is close to the optimal value among all scalar linear MDS codes with the optimal repair property.

\subsection{Organization of the paper}
In Section~\ref{sect:warmup} below, we present a relatively simple construction of RS codes that achieves the cut-set bound for the repair of any two erasures. This construction contains the main ideas of the later part and hopefully makes it easier to understand
the case of an arbitrary number of erasures. In Section~\ref{Sect:uee}, we
present our main construction of RS codes that achieve the cut-set bound for the repair of any number of failed nodes from any set of helper nodes.

\section{Optimal repair of two erasures}\label{sect:warmup}
In this section we present an explicit construction of RS codes that achieve the cut-set bound \eqref{eq:cs1} for the repair of any two failed nodes.
\subsection{Some definitions}
Let us first recall some basic concepts that will be used throughout the paper.

\begin{definition}[Dual code] Let $\cC$ be a linear code of length $n$ over a finite field $\mathbb{K}$. 
The dual code of $\cC$ is the linear subspace of $\mathbb{K}^n$ defined by
$$
\cC^{\perp}=\big\{x=(x_1,\dots,x_n) \in \mathbb{K}^n \big|\sum_{i=1}^n x_i c_i = 0 \quad \forall c=(c_1,\dots c_n)\in\cC \big\}.
$$
\end{definition}

\begin{definition}
A \emph{generalized Reed-Solomon code} $\text{\rm GRS}_{\mathbb{K}}(n,k,\Omega,v)\subseteq \mathbb{K}^n$ of dimension $k$ over $\mathbb{K}$ 
with evaluation points $\Omega=\{\omega_1,\omega_2,\dots,\omega_n\}\subseteq \mathbb{K}$  is the set of vectors
\begin{align*}
\{(v_1f(\omega_1),\dots,v_nf(\omega_n))\in \mathbb{K}^n:f\in \mathbb{K}[x], \deg f\le k-1\},
\end{align*}
where $v=(v_1,\dots,v_n)\in (\mathbb{K}^\ast)^n$ are some nonzero elements. If $v=(1,\dots,1),$ then the GRS code is called
a \emph{Reed-Solomon code} and is denoted as $\text{\rm RS}_{\mathbb{K}}(n,k,\Omega)$.
\end{definition}

It is well known \cite[p.304]{Macwilliams77} that 
   \begin{equation}\label{eq:grs}
    (\text{\rm RS}_{\mathbb{K}}(n,k,\Omega))^\bot=\text{\rm GRS}_{\mathbb{K}}(n,n-k,\Omega,v),
  \end{equation}
where $v_i=\prod_{j\ne i}(\omega_i-\omega_j)^{-1}, i=1,\dots,n$ (the dual of an RS code is a GRS code). 
   
Let $E$ be the extension of degree $t$ of a finite field $F=\ff_q$. The trace $\trace_{E/F}$ is a mapping from $E$ to $F$
defined as
   $$
   \trace_{E/F}(x)=\sum_{i=0}^{t-1} x^{q^i}.
   $$
The trace has the following \emph{transitivity property}: let $K$ be a finite algebraic extension of $E$, then for all $a\in K,$
   \begin{equation}\label{eq:trans}
    \trace_{K/F}(a)=\trace_{E/F}(\trace_{K/E}(a)).
   \end{equation}

\subsection{Code construction}
Let us fix the values of the code length $n$ and dimension $k.$ Let $d, k\le d\le n-2$ be the number of helper nodes used for recovery.  In the case of $h=2$ the cut-set bound \eqref{eq:cutset} has the form $\beta(2,d)=\frac{2dl}{d+2-k}.$
Our goal will be accomplished if we construct codes and a repair procedure that relies on downloading a ${2}/(d+2-k)$ fraction of the node contents from each of the helper nodes.

Let $\mathbb{F}_p$ be a finite field (for simplicity we can take $p=2$). 
Define $s=s_1s_2,$ where
\begin{equation}\label{eq:s12}
s_1=d+1-k, \quad s_2=d+2-k.
\end{equation}
Let $p_1,\dots,p_n$ be $n$ distinct primes such that
\begin{equation}\label{eq:pms}
p_i\equiv 1 \;\text{mod}\, s \;\;\text{~for all~} i=1,2,\dots,n.
\end{equation}
According to Dirichlet's theorem, there are infinitely many such primes. For $i=1,\dots,n$, let $\alpha_i$ be an element of 
degree $p_i$ over $\mathbb{F}_p$, i.e., $[\mathbb{F}_p(\alpha_i):\mathbb{F}_p]=p_i$, and define 
\begin{equation}\label{eq:defF}
\mathbb{F}:=\mathbb{F}_p(\alpha_1,\dots,\alpha_n).
\end{equation}
Note that for any subset of indices $A\subseteq [n]$, the field $\mathbb{F}_p(\{\alpha_i:i\in A\})$ is an extension of $\mathbb{F}_p$
of degree $\prod_{i\in A}p_i,$ and in particular, $\mathbb{F}$ has degree $\prod_{i=1}^np_i$ over $\mathbb{F}_p$. 

Finally, let $\mathbb{K}$ be an algebraic extension of $\mathbb{F}$ of degree $s$ and let $\beta\in \mathbb{K}$ 
be such that %of degree $s$ over $\mathbb{F}$ such that 
\begin{equation}\label{eq:bbk}
\mathbb{K}=\mathbb{F}(\beta)
\end{equation}
($\beta$ always exists by the primitive element theorem).

The codes that we construct have length $n$ and use $\{\alpha_1,\dots,\alpha_n\}$ as the set of evaluation points.
Our results are summarized in the following theorem.
\begin{theorem} \label{thm:2err} 
Let $k,n,d$ be any positive integers such that $k< d < n.$ Let $\Omega=\{\alpha_1,\dots,\alpha_n\}$, where $\alpha_i,i=1,\dots,n$ is an element of degree $p_i$ over $\mathbb{F}_p$ and $p_i$ is the $i$th smallest prime that satisfies \eqref{eq:pms}.
 Then the code $\cC:=\text{\rm RS}_{\mathbb{K}}(n,k,\Omega)$ 
 %achieves the cut-set bound for the repair of any two nodes from any $d$ helper nodes. In other words, $\cC$ 
 has the $(2,d)$-optimal repair property. 

The sub-packetization value of the code $\cC$ equals
   \begin{equation}\label{eq:2sp}
   l=[\mathbb{K}:\mathbb{F}_p]=s \prod_{i=1}^n p_i.
   \end{equation}
   For fixed $r$ and growing $n$ we have $l=e^{(1+o(1))n \log n}.$
 \end{theorem}
\begin{IEEEproof}
We write a codeword of $\cC$ as $(c_1,\dots,c_n)$. Referring to \eqref{eq:cutset}, let $\cF=\{i_1,i_2\}$ be the indices
of the failed nodes, and let $\cR\subseteq [n]\backslash\{i_1,i_2\}$ be the set of $d$ helper nodes used in repair.
Our repair scheme is performed over the field 
\begin{equation}\label{eq:DFF}
F:=\mathbb{F}_p(\{\alpha_j:j\in [n]\setminus \{i_1,i_2\}\}).
\end{equation}
It is clear that $\mathbb{F}=F(\alpha_{i_1},\alpha_{i_2})$ and $[\mathbb{F}:F]=p_{i_1} p_{i_2}$.
As a consequence, $[\mathbb{K}:F]=s p_{i_1} p_{i_2}$.
Our strategy is as follows: 
\vspace{.1in}\begin{enumerate}
  \item[$(i)$] First repair node $c_{i_1}$ from the helper nodes in $\cR$. We show that this can be done by downloading 
$({s p_{i_1} p_{i_2}})/s_1$ symbols of $F$ from each of the helper nodes in $\cR$.
\vspace{.05in}  
  \item[$(ii)$] 
Then we use the helper nodes in $\cR$ together with the already repaired node $c_{i_1}$ to repair the node $c_{i_2}$, and we show that this can be done by downloading 
$\frac{s p_{i_1} p_{i_2}}{s_2}$ symbols of $F$ from each of the helper nodes in $\cR$.
\vspace{.05in} 
   \item[$(iii)$] We show that for each helper node in $\cR$, the two sets of downloaded symbols (for the repair of $c_{i_1}$ and $c_{i_2},$ respectively) have an overlap of size $p_{i_1} p_{i_2}$.
 \end{enumerate}  
\vspace{.1in} Therefore in total we need to download 
  \begin{align*}
  s_2p_{i_1} p_{i_2}&+s_1p_{i_1} p_{i_2}-p_{i_1} p_{i_2}\\
     &=2s_1p_{i_1} p_{i_2}\\
     &=\frac 2{s_2} s p_{i_1} p_{i_2}
  \end{align*}
   symbols of $F$ from each of the helper nodes. This forms a $2/(d+2-k)$ proportion of the node contents, and so the scheme achieves the cut-set bound \eqref{eq:cs1} with equality.

%More specifically, we define two constants
%\begin{equation}\label{eq:s12}
%s_1=d+1-k, \quad s_2=d+2-k.
%\end{equation}
%Notice that $s=s_1s_2$.

\vspace*{.1in} Proceeding with the implementation of the above plan, define the sets $W_{i_1}, W_{i_1}^{(1)}, W_{i_1}^{(2)}$
and $W_{i_2}, W_{i_2}^{(1)}, W_{i_2}^{(2)}$ as follows:
\begin{equation}\label{eq:wi}
\begin{aligned}
W_{i_1}^{(1)} := &
\Big\{ \beta^{u_1} \alpha_{i_1}^{u_1 + qs_1}  :
u_1=0,1,\dots,s_1-1; 
 q=0,1,\dots,\frac{p_{i_1}-1}{s_1}-1 \Big\}, \\
W_{i_1}^{(2)} := &
\Big\{  \alpha_{i_1}^{p_{i_1}-1} \sum_{u_1=0}^{s_1-1} \beta^{u_1} \Big\}, \\
W_{i_1} := & W_{i_1}^{(1)} \cup W_{i_1}^{(2)}; \\[.1in]
W_{i_2}^{(1)} := &
\Big\{ \beta^{u_2 s_1} \alpha_{i_2}^{u_2 + qs_2}  :
u_2=0,1,\dots,s_2-1; 
 q=0,1,\dots,\frac{p_{i_2}-1}{s_2}-1 \Big\}, \\
W_{i_2}^{(2)} := &
\Big\{   \alpha_{i_2}^{p_{i_2}-1}\sum_{u_2=0}^{s_2-1} \beta^{u_2 s_1} \Big\}, \\
W_{i_2} := & W_{i_2}^{(1)} \cup W_{i_2}^{(2)}.
\end{aligned}
\end{equation}

We further define two sets of elements
\begin{equation}\label{eq:defSi}
S_{i_1} := \bigcup_{u_2=0}^{s_2-1} \bigcup_{q_2=0}^{p_{i_2}-1}
\Big( \beta^{u_2 s_1} \alpha_{i_2}^{q_2} W_{i_1} \Big), \quad
S_{i_2} := \bigcup_{u_1=0}^{s_1-1} \bigcup_{q_1=0}^{p_{i_1}-1}
\Big( \beta^{u_1} \alpha_{i_1}^{q_1}  W_{i_2}\Big),
\end{equation}
where the product of an element $\alpha$ and a set $S$ is defined as the set $\alpha S = \{\gamma \alpha: \gamma \in S\}$.
It is clear that $|S_{i_1}|=s_2 p_{i_1} p_{i_2}$ and $|S_{i_2}|=s_1 p_{i_1} p_{i_2}$.

The theorem will follow from the next three lemmas.

\begin{lemma} \label{lem:S1}
Node $c_{i_1}$ can be repaired from the set of symbols 
$\{\trace_{\mathbb{K}/F}(\gamma v_jc_j):\gamma\in S_{i_1},j\in \cR\}$.
\end{lemma}

\begin{lemma} \label{lem:S2}
Node $c_{i_2}$ can be repaired from $c_{i_1}$ together with the set of symbols 
$\{\trace_{\mathbb{K}/F}(\gamma v_jc_j):\gamma\in S_{i_2},j\in \cR\}$.
\end{lemma}

For a vector space $V$ over a field $F$ and a set of vectors $A \subset V$, let $\spun_F(A)$ be the linear span of $A$ over $F$.

\begin{lemma} \label{lem:ints}
$$
\dim_F(\spun_F(S_{i_1}) \cap \spun_F(S_{i_2})) = p_{i_1} p_{i_2}.
$$
\end{lemma}

Let us first show that these three lemmas indeed imply Theorem~\ref{thm:2err}. 
On account of Lemmas \ref{lem:S1} and \ref{lem:S2} the sets of symbols
   $$
   D_j=\{\trace_{\mathbb{K}/F}(\gamma v_jc_j):\gamma\in S_{i_1}\cup S_{i_2}\}, \quad j\in \cR
   $$ 
suffice to find the values $c_{i_1}$ and $c_{i_2}.$
In their turn, the elements in the set $D_j, j\in \cR$ will be found once we download the elements in the set
$\{\trace_{\mathbb{K}/F}(\gamma v_jc_j):\gamma\in B\}$, where
 the elements in $B$ form a basis of 
$\spun_F(S_{i_1}) + \spun_F(S_{i_2})$ over $F$.
Therefore the number of symbols in $F$ that we need to download from each helper node is equal to the dimension of
$\spun_F(S_{i_1}) + \spun_F(S_{i_2})$ over $F$. We have
   \begin{equation}\label{eq:ios}
\dim_F(\spun_F(S_{i_1}) + \spun_F(S_{i_2}))= |S_{i_1}| + |S_{i_2}| - \dim_F(\spun_F(S_{i_1}) \cap \spun_F(S_{i_2})).
   \end{equation}
Using Lemma~\ref{lem:ints}, we now obtain
  $$
 \dim_F(\spun_F(S_{i_1}) + \spun_F(S_{i_2})) 
=  2 s_1 p_{i_1} p_{i_2} = \frac{2}{d+2-k} s p_{i_1} p_{i_2}.
  $$
Since $[\mathbb{K}:F]=s p_{i_1} p_{i_2}$, we conclude that the repair bandwidth of $\{c_{i_1},c_{i_2}\}$ from the helper nodes $\{c_j:j\in\cR\}$ indeed achieves the cut-set bound \eqref{eq:cs1}. 

Moreover, since the repair field of the pair $\{i_1,i_2\}$ is $\mathbb{F}_p(\{\alpha_j:j\in [n]\setminus \{i_1,i_2\}\})$, the largest common repair field for all possible pair of coordinates is ${\mathbb F}_p.$ This justifies the claim about the sub-packetization of our construction
made in \eqref{eq:2sp}.
\end{IEEEproof}

\vspace*{.1in}
Next we prove Lemmas~\ref{lem:S1}-\ref{lem:ints}.

\vspace*{.1in}{\em Proof of Lemma~\ref{lem:S1}:}
The proof of this lemma is an extension of the argument of Theorem 4 in \cite{Tamo17RS} (more on this in Remark \ref{remark2} in the end of this section).
Define a field 
\begin{equation}\label{eq:fi1}
F_{i_1}:=\mathbb{F}_p(\{\alpha_j:j\neq i_1\}).
\end{equation}
According to \eqref{eq:defF}, we have
\begin{equation}\label{eq:rFi1}
\mathbb{F}= F_{i_1}(\alpha_{i_1}), \text{~and~} [\mathbb{F} : F_{i_1}]=p_{i_1}. 
\end{equation} 
Let $h_1(x)$ be the annihilator polynomial of the set $\{\alpha_j: j\in[n]\setminus (\cR\cup\{i_1\}) \}$, i.e., 
\begin{equation}\label{eq:defh1}
h_1(x)=\prod_{j\in[n]\setminus (\cR\cup\{i_1\})}(x-\alpha_j).
\end{equation}
As  remarked above \eqref{eq:grs}, the dual code of $\cC$ is $\cC^\bot=\text{\rm GRS}_{\mathbb{K}}(n,n-k,\Omega,v),$ where  $v=(v_1,\dots,v_n) \in (\mathbb{K}^*)^n.$ 
Clearly, $\deg(x^t h_1(x))\leq s_1-1+n-(d+1)<n-k$ for all $t=0,1,\dots,s_1-1,$ so for any such $t$ we have
\begin{equation}\label{eq:dual}
(v_1\alpha_1^t h_1(\alpha_1),\dots,v_n \alpha_n^t h_1(\alpha_n))\in\cC^\bot.
\end{equation}
These $s_1$ dual codewords will  be used to recover the $i_1$-th coordinate. 
We define a set $T_{i_1}$ as follows:
\begin{equation}\label{eq:defT1}
T_{i_1} :=  \bigcup_{u_2=0}^{s_2-1} 
\Big( W_{i_1} \beta^{u_2 s_1} \Big).
\end{equation}
The elements in $T_{i_1}$ will also be used to recover the $i_1$-th coordinate.
Using \eqref{eq:defSi}, it is easy to verify the following relation:
\begin{equation}\label{eq:SRT}
S_{i_1}  = \bigcup_{q_2=0}^{p_{i_2}-1} T_{i_1} \alpha_{i_2}^{q_2}.
\end{equation}

Let $c=(c_1,\dots,c_n)\in \cC$ be a codeword, and let us construct a repair scheme for the coordinate (node) $c_i$ using
the values $\{c_j:j\in \cR\}$. Rewrite \eqref{eq:dual} as follows:
$$
\sum_{j=1}^n  v_j\alpha_j^t h_1(\alpha_j) c_j =0, \quad  t=0,\dots,s_1-1.
$$
As an immediate consequence, for all $t=0,\dots,s_1-1$ and $\gamma \in T_{i_1},$ we have
\begin{equation}\label{eq:inter}
\sum_{j=1}^n \trace_{\mathbb{K}/F_{i_1}} (\gamma v_j\alpha_j^t h_1(\alpha_j) c_j) =0 .
\end{equation}
Let us write \eqref{eq:inter} in the following form:
\begin{equation}\label{eq:rcv}
\begin{aligned}
\trace_{\mathbb{K}/F_{i_1}} (\gamma \alpha_{i_1}^t v_{i_1} h_1(\alpha_{i_1})c_{i_1}) & 
= - \sum_{j\neq i_1} \trace_{\mathbb{K}/F_{i_1}} (\gamma v_j \alpha_j^t h_1(\alpha_j)c_j) \\
& = - \sum_{j\in \cR} \trace_{\mathbb{K}/F_{i_1}} (\gamma v_j \alpha_j^t h_1(\alpha_j)c_j) \\
& = - \sum_{j\in \cR} \alpha_j^t h_1(\alpha_j) \trace_{\mathbb{K}/F_{i_1}} (\gamma v_j  c_j)
\text{~for all~} t=0,\dots,s_1-1 \text{~and all~} \gamma \in T_{i_1},
\end{aligned}
\end{equation}
where the second equality follows from \eqref{eq:defh1} and the third follows from the fact that the trace mapping $\trace_{\mathbb{K}/F_{i_1}}$ is $F_{i_1}$-linear, and that $\alpha_j\in F_{i_1}$
and $h_1(\alpha_j) \in F_{i_1}$ for all $j \neq i_1$.

Next we observe that the set $\{\gamma\alpha_{i_1}^t:  t=0,1,\dots,s_1-1 ;\, \gamma\in T_{i_1}\}$ of size $sp_{i_1}$ forms a basis of  $\mathbb{K}$ over  $F_{i_1}$ (see Prop.~\ref{prop:sumt} in Appendix~\ref{ap:PropT}). Since  $v_{i_1} h_1(\alpha_{i_1})\neq 0,$ the set
$\{\gamma\alpha_{i_1}^t v_{i_1} h_1(\alpha_{i_1}):  t=0,1,\dots,s_1-1 ;\,  \gamma\in T_{i_1}\}$  also forms a basis.
Therefore, the value of $c_{i_1}$ can be calculated from the set
  $$
  \{\trace_{\mathbb{K}/F_{i_1}} (\gamma \alpha_{i_1}^t v_{i_1} h_1(\alpha_{i_1})c_{i_1})
:t=0,1,\dots,s_1-1 ;\,  \gamma\in T_{i_1} \}.
 $$

Using \eqref{eq:rcv}, we conclude that the value of $c_{i_1}$ can be calculated from
$\{\trace_{\mathbb{K}/F_{i_1}} (\gamma v_j  c_j):\gamma\in T_{i_1}, j\in\cR\}$.
To complete the proof of Lemma~\ref{lem:S1}, it suffices to show that the elements in the set
$\{\trace_{\mathbb{K}/F_{i_1}} (\gamma v_j  c_j):\gamma\in T_{i_1}, j\in\cR\}$
can be calculated from 
$\{\trace_{\mathbb{K}/F}(\gamma v_jc_j):\gamma\in S_{i_1},j\in \cR\}$. This is an immediate consequence of equation \eqref{eq:SRT}.
Indeed, observe that $F_{i_1}=F(\alpha_{i_2})$ and that $\{1,\alpha_{i_2},\dots,\alpha_{i_2}^{p_{i_2}-1}\}$ forms a basis of $F_{i_1}$ over $F$. Therefore, for every $\gamma\in T_{i_1}$ and every $j\in\cR$, the value of $\trace_{\mathbb{K}/F_{i_1}} (\gamma v_j  c_j)$ can be calculated from
$\{\trace_{F_{i_1}/F}(\trace_{\mathbb{K}/F_{i_1}} (\gamma v_j  c_j) \alpha_{i_2}^{q_2}): q_2=0,1,\dots,p_{i_2}-1\}$.
Observe that
$$
\trace_{F_{i_1}/F}(\trace_{\mathbb{K}/F_{i_1}} (\gamma v_j  c_j) \alpha_{i_2}^{q_2} )
= \trace_{F_{i_1}/F}(\trace_{\mathbb{K}/F_{i_1}} (\gamma v_j  c_j \alpha_{i_2}^{q_2}) )
= \trace_{\mathbb{K}/F} (\gamma v_j  c_j \alpha_{i_2}^{q_2}),
$$
where the first equality follows from the fact that $\alpha_{i_2}\in F_{i_1}$, and the second equality follows from \eqref{eq:trans}.
Therefore, for every $\gamma\in T_{i_1}$ and every $j\in\cR$, the value of $\trace_{\mathbb{K}/F_{i_1}} (\gamma v_j  c_j)$ can be calculated from
$\{\trace_{\mathbb{K}/F} (\gamma v_j  c_j \alpha_{i_2}^{q_2}):q_2=0,1,\dots,p_{i_2}-1\}
\subseteq \{\trace_{\mathbb{K}/F}(\gamma v_jc_j):\gamma\in S_{i_1},j\in \cR\}$, where the inclusion follows from \eqref{eq:SRT}.
Therefore we have shown that the elements in the set
$\{\trace_{\mathbb{K}/F_{i_1}} (\gamma v_j  c_j):\gamma\in T_{i_1}, j\in\cR\}$
can be calculated from 
$\{\trace_{\mathbb{K}/F}(\gamma v_jc_j):\gamma\in S_{i_1},j\in \cR\}$, and this completes the proof of Lemma~\ref{lem:S1}. \hfill $\blacksquare$

\vspace*{.1in}{\em Proof of Lemma~\ref{lem:S2}:}
Let $h_2(x)$ be the annihilator polynomial of the set $\{\alpha_j: j\in[n]\setminus (\cR\cup\{i_1,i_2\}) \}$, i.e., 
\begin{equation}\label{eq:defh2}
h_2(x)=\prod_{j\in[n]\setminus (\cR\cup\{i_1,i_2\})}(x-\alpha_j).
\end{equation}
Clearly, $\deg(x^t h_2(x))\leq s_2-1+n-(d+2)<n-k$ for all $t=0,1,\dots,s_2-1,$ so for any such $t$ we have
\begin{equation}\label{eq:dual2}
(v_1\alpha_1^t h_2(\alpha_1),\dots,v_n \alpha_n^t h_2(\alpha_n))\in\cC^\bot.
\end{equation}
These $s_2$ dual codewords will  be used to recover the $i_2$-th coordinate. 
Let us construct a repair scheme for the coordinate (node) $c_{i_2}$ using
the values $\{c_j:j\in \cR\cup\{i_1\}\}$. Rewrite \eqref{eq:dual2} as follows:
$$
\sum_{j=1}^n  v_j\alpha_j^t h_2(\alpha_j) c_j =0 \text{~for all~}  t=0,\dots,s_2-1.
$$
Computing the trace, we obtain
\begin{equation}\label{eq:inter2}
\sum_{j=1}^n \trace_{\mathbb{K}/F} (\gamma v_j\alpha_j^t h_2(\alpha_j) c_j) =0 \text{~for all~}  t=0,\dots,s_2-1 \text{~and all~} \gamma \in S_{i_2}.
\end{equation}
Let us write \eqref{eq:inter2} in the following form:
\begin{equation}\label{eq:rcv2}
\begin{aligned}
\trace_{\mathbb{K}/F} (\gamma \alpha_{i_2}^t v_{i_2} h_2(\alpha_{i_2})c_{i_2}) & 
= - \sum_{j\neq i_2} \trace_{\mathbb{K}/F} (\gamma v_j \alpha_j^t h_2(\alpha_j)c_j) \\
& = - \trace_{\mathbb{K}/F} (\gamma v_{i_1} \alpha_{i_1}^t h_2(\alpha_{i_1})c_{i_1})
- \sum_{j\in \cR} \trace_{\mathbb{K}/F} (\gamma v_j \alpha_j^t h_2(\alpha_j)c_j) \\
& = - \trace_{\mathbb{K}/F} (\gamma v_{i_1} \alpha_{i_1}^t h_2(\alpha_{i_1})c_{i_1})
 - \sum_{j\in \cR} \alpha_j^t h_2(\alpha_j) \trace_{\mathbb{K}/F} (\gamma v_j  c_j)\\
& \hspace*{0.4in} \text{~for all~} t=0,\dots,s_2-1 \text{~and all~} \gamma \in S_{i_2},
\end{aligned}
\end{equation}
where the second equality follows from \eqref{eq:defh2} and the third follows from the fact that the trace mapping $\trace_{\mathbb{K}/F}$ is $F$-linear, and that $\alpha_j\in F$ and $h_2(\alpha_j) \in F$ for all $j\in \cR$.

According to Prop.~\ref{prop:sumS} in Appendix~\ref{ap:PropS}, the set $\{\gamma\alpha_{i_2}^t:  t=0,1,\dots,s_2-1 ;\, \gamma\in S_{i_2}\}$ forms a basis of  $\mathbb{K}$ over  $F$
and so does the set  $\{\gamma\alpha_{i_2}^t v_{i_2} h_2(\alpha_{i_2}):  t=0,1,\dots,s_2-1 ;\,  \gamma\in S_{i_2}\}$ (recall that $v_{i_2} h_2(\alpha_{i_2})\neq 0$).
Hence the value of $c_{i_2}$ can be calculated from 
$\{\trace_{\mathbb{K}/F} (\gamma \alpha_{i_2}^t v_{i_2} h_2(\alpha_{i_2})c_{i_2})
:t=0,1,\dots,s_2-1 ;\,  \gamma\in S_{i_2} \}$.

Using \eqref{eq:rcv2}, we conclude that the value of $c_{i_2}$ can be calculated from
the value of $c_{i_1}$ and the values of elements in the set
$\{\trace_{\mathbb{K}/F} (\gamma v_j  c_j):\gamma\in S_{i_2}, j\in\cR\}$.
This completes the proof of Lemma~\ref{lem:S2}.
\hfill $\blacksquare$

\vspace*{.1in}{\em Proof of Lemma~\ref{lem:ints}:}
Using the cut-set bound on the left-hand side of Equation \eqref{eq:ios}, we obtain the inequality
    $$
\dim_F(\spun_F(S_{i_1}) \cap \spun_F(S_{i_2})) \le p_{i_1} p_{i_2}.
    $$
Let us prove that
     \begin{equation}\label{eq:isd}
\dim_F(\spun_F(S_{i_1}) \cap \spun_F(S_{i_2})) \ge p_{i_1} p_{i_2}.
     \end{equation}
To this end, we will find $p_{i_1} p_{i_2}$ elements in
$\spun_F(S_{i_1}) \cap \spun_F(S_{i_2})$ that are linearly independent over $F$.

Let us recall the definitions of $W_{i_1}$ and $W_{i_2}$ given in \eqref{eq:wi}.
Note that
$$
W_{i_2}  \subseteq \spun_F \Big(\bigcup_{u_2=0}^{s_2-1} \bigcup_{q_2=0}^{p_{i_2}-1} \{\beta^{u_2 s_1} \alpha_{i_2}^{q_2}\}\Big).
$$
Combining this with \eqref{eq:defSi}, we deduce that
$$
W_{i_1}\odot W_{i_2} \subseteq W_{i_1} \odot
\spun_F\Big(\bigcup_{u_2=0}^{s_2-1} \bigcup_{q_2=0}^{p_{i_2}-1} \{\beta^{u_2 s_1} \alpha_{i_2}^{q_2}\}\Big)
\subseteq \spun_F(S_{i_1}),
$$
where the product $\odot$ of sets $A_1$ and $A_2$ is defined as 
\begin{equation}\label{eq:dod}
A_1\odot A_2 
:=\{\gamma_1 \gamma_2:\gamma_1\in A_1, \gamma_2\in A_2\}.
\end{equation}
Similarly, we also have $W_{i_1}\odot W_{i_2} \subseteq \spun_F(S_{i_2})$, and therefore 
  \begin{equation}\label{eq:o.}
W_{i_1}\odot W_{i_2} \subseteq ( \spun_F(S_{i_1}) \cap \spun_F(S_{i_2}) ).
  \end{equation}
It is clear that 
$|W_{i_1}\odot W_{i_2}| = |W_{i_1}| |W_{i_2}| = p_{i_1} p_{i_2}$.
Moreover, for every $u\in\{0,1,\dots,s-1\}$, every $q_1\in\{0,1,\dots,p_{i_1}-1\}$ and every
$q_2\in\{0,1,\dots,p_{i_2}-1\}$, $\beta^u \alpha_{i_1}^{q_1} \alpha_{i_2}^{q_2}$
appears at most once\footnote{Such an element may be itself contained in $W_{i_1}\odot W_{i_2},$ or appear as a summand of an element in $W_{i_1}\odot W_{i_2}$} in $W_{i_1}\odot W_{i_2}$.
Since the elements in the set
$\{\beta^u \alpha_{i_1}^{q_1} \alpha_{i_2}^{q_2}: u=0,1,\dots,s-1; q_1=0,1,\dots,p_{i_1}-1;
q_2=0,1,\dots,p_{i_2}-1\}$ are linearly independent over $F$,
we deduce that all the elements in
$W_{i_1}\odot W_{i_2}$ are linearly independent over $F$.
Now \eqref{eq:isd} follows from \eqref{eq:o.}, and this completes the proof of Lemma~\ref{lem:ints}.
\hfill $\blacksquare$

\begin{remark}{\rm 
It is obvious from the proofs that the code construction in this section also has the $(1,d)$-optimal repair property and $(1,d+1)$-optimal repair property. In other words, the repair of any single erasure from any $d$ or $d+1$ helper nodes also achieves the cut-set bound.}
\end{remark}

\begin{remark} \label{remark2}
{\rm Let us point out some new ingredients in the repair of multiple erasures compared to the repair of a single erasure 
\cite{Tamo17RS}. These ideas will be used in the next section where we present a scheme for repairing an arbitrary number of erasures.

The first one appears in the proof of Lemma~\ref{lem:S1}. The proof of Lemma~\ref{lem:S1} consists of two parts: in the first part we show that $c_{i_1}$ can be calculated from $\{\trace_{\mathbb{K}/F_{i_1}} (\gamma v_j  c_j):\gamma\in T_{i_1}, j\in\cR\}$;
in the second part we show that the elements in the set $\{\trace_{\mathbb{K}/F_{i_1}} (\gamma v_j  c_j):\gamma\in T_{i_1}, j\in\cR\}$ can be calculated from $\{\trace_{\mathbb{K}/F}(\gamma v_jc_j):\gamma\in S_{i_1},j\in \cR\}$.
The proof of the first part is the same as the proof of Theorem 4 in \cite{Tamo17RS}, and the new idea lies in the second part, where in particular we use transitivity of the trace mapping.

The other new ingredient is Lemma~\ref{lem:ints}, where we calculate the dimension of the intersection. Similar calculations also allow us to achieve the cut-set bound for the repair of more than two erasures in the next section. }
\end{remark}

\begin{remark}\label{remark3}
{\rm Finally, consider the full subfield lattice ordered by inclusion, starting with the field $\ff_p$ as the root and ending
with $\ff$ as the unique maximal element, i.e., the subset lattice of the $n$-set 
$\{\alpha_1,\alpha_2,\dots,\alpha_n\}$. In the above repair scheme we relied on subfields of the form $F$
(see \eqref{eq:DFF}), i.e., those that
contain all but two elements of this set. In a similar way, in our repair scheme for $h\ge 2$ erasures below we rely on subfields that contain $n-h$ of the $n$ elements of the set $\{\alpha_1,\alpha_2,\dots,\alpha_n\}$.
}
\end{remark}

\section{Universally achieving cut-set bound for any number of erasures}\label{Sect:uee}
In this section we present an explicit construction of $(n,k=n-r)$ RS codes with the universal $(h,d)$-optimal repair property for all $h\le r$ and all $k\le d \le n-h$ simultaneously. In other words, the constructed codes can repair any number of erasures from any set of helper nodes with repair bandwidth achieving the cut-set bound.
Even though the notation in this section is somewhat more involved than above, the main ideas are similar to the ideas used in the construction of RS codes with optimal repair for two erasures.

We again begin with a finite field $\mathbb{F}_p$ (for simplicity we can take $p=2$). Let $p_1,\dots,p_n$ be $n$ distinct primes such that
\begin{equation}\label{eq:pmr}
p_i\equiv 1 \;\text{mod}\, r! \;\;\text{~for all~} i=1,2,\dots,n.
\end{equation}
According to Dirichlet's theorem, there are infinitely many such primes. For $i=1,\dots,n$, let $\alpha_i$ be an element of 
degree $p_i$ over $\mathbb{F}_p$, i.e., $[\mathbb{F}_p(\alpha_i):\mathbb{F}_p]=p_i$, and define 
\begin{equation}\label{eq:defFr}
\mathbb{F}:=\mathbb{F}_p(\alpha_1,\dots,\alpha_n).
\end{equation}
Note that for any subset of indices $A\subseteq [n]$, the field $\mathbb{F}_p(\{\alpha_i:i\in A\})$ is an extension of $\mathbb{F}_p$
of degree $\prod_{i\in A}p_i,$ and in particular, $\mathbb{F}$ has degree $\prod_{i=1}^np_i$ over $\mathbb{F}_p$. 

Let $\mathbb{K}$ be an algebraic extension of $\mathbb{F}$ of degree $r!$
and let $\beta\in \mathbb{K}$ be an element of degree $r!$ over $\mathbb{F}$ such that 
     \begin{equation}\label{eq:rst}
\mathbb{K}=\mathbb{F}(\beta).
     \end{equation}

Similarly to \eqref{eq:s12}, we define the following $h$ constants: for $i=1,2,\dots,h$, let
\begin{equation}\label{eq:sh}
s_i=d+i-k.
\end{equation}
Note that $s_i\le r$ for all $i\le h,$ and so $s_i|(p_i-1)$. It will also be convenient to have a notation for partial products of the numbers $s_i$. Namely, let
\begin{equation}\label{eq:tp}
t_1=1;\quad t_i=\prod_{j=1}^{i-1}s_j ,\; i=2,3,\dots,h+1
\end{equation}
and let 
    \begin{equation}\label{eq:sh+1}  
      s_{h+1}: = \frac{r!}{t_{h+1}}.
    \end{equation}
Observe the following simple facts:
\begin{align}
\Big\{\sum_{i=1}^h u_i t_i :
u_i=0,1,\dots,s_i-1; i=1,2,\dots,h\Big\} 
& =  \{0,1,2,\dots, t_{h+1} - 1\}, \nonumber \\ 
\Big\{\sum_{i=1}^{h+1} u_i t_i: 
u_i=0,1,\dots,s_i-1 \text{~for all~} i=1,2,\dots,h+1\Big\} 
& =  \{0,1,2,\dots, r! - 1\}. \label{eq:nsu}
\end{align}

Our construction of codes with the universal $(h,d)$ optimal repair property relies on RS codes with evaluation points
$\alpha_1,\dots,\alpha_n.$ 
Specifically, the following is true:

\begin{theorem} \label{thm:ue} 
Let $k,n$ be any positive integers such that $k < n$ and let $p_i,i=1,2,\dots,n$ be the $i$th smallest prime that 
satisfies \eqref{eq:pmr}. Let $\Omega=\{\alpha_1,\dots,\alpha_n\}$, where $\alpha_i,i=1,\dots,n$ is an element of degree $p_i$ over $\mathbb{F}_p.$
 The code $\cC:=\text{\rm RS}_{\mathbb{K}}(n,k,\Omega)$ achieves the cut-set bound for the repair of any number $h$ of failed nodes from any set of $d$ helper nodes provided that $h\le r$ and $k\le d \le n-h.$ 
In other words, $\cC$ has the universal $(h,d)$-optimal repair property for all $h$ and $d$ simultaneously.

The sub-packetization value of the code $\cC$ equals
   \begin{equation}\label{eq:sp}
   l=[\mathbb{K}:\mathbb{F}_p]=r!\prod_{i=1}^n p_i.
   \end{equation}
   For fixed $r$ and growing $n$ we have $l=e^{(1+o(1))n \log n}.$
 \end{theorem}

\begin{IEEEproof}
We write a codeword of $\cC$ as $(c_1,\dots,c_n)$.
Suppose that the number of failed nodes is $h$ and the number of helper nodes is $d$ for some $h\le r$ and some $k\le d \le n-h$.
Without loss of generality, we assume that
the indices of the failed nodes are $\cF=\{1,2,\dots,h\}$ and the indices of helper nodes
are $\cR=\{h+1,h+2,\dots,h+d\}$. 
Our repair scheme of these $h$ failed nodes is performed over the field
$$
F_{[h]}:=\mathbb{F}_p(\{\alpha_i:i\in [n]\setminus [h]\})
$$
(recall that $[h]:=\{1,2,\dots,h\}$; see also Remark~\ref{remark3}).
It is clear that $\mathbb{F}=F_{[h]}(\alpha_1,\alpha_2,\dots,\alpha_h)$ and $[\mathbb{F}:F_{[h]}]=\prod_{i=1}^h p_i$.
As a consequence, 
   \begin{equation}\label{eq:dim}
   [\mathbb{K}:F_{[h]}]=r! \prod_{i=1}^h p_i.
   \end{equation}
Our strategy is as follows: 
   \begin{enumerate}
  \item[$(i)$]
Begin with repairing node $c_1$ from the helper nodes in $\cR$. We show that this can be done by downloading 
$\frac{r! \prod_{i=1}^h p_i}{d+1-k}$ symbols of $F_{[h]}$ from each of the helper nodes in $\cR$.
\item[$(ii)$]
Then we use the helper nodes in $\cR$ together with the already repaired node $c_1$ to repair the node $c_2$, and we show that this can be done by downloading 
$\frac{r! \prod_{i=1}^h p_i}{d+2-k}$ symbols of $F_{[h]}$ from each of the helper nodes in $\cR$.
\item[$(iii)$]
We continue in this way until we use the helper nodes in $\cR$ together with the already repaired nodes $c_1,c_2,\dots,c_{h-1}$ to repair $c_h$.
\item[$(iv)$]
Finally we show that for each helper node in $\cR,$ the $h$ sets of downloaded symbols (for the repair of $c_1,c_2,\dots,c_h$ respectively) have overlaps, and that after removing the overlapping parts it suffices to download 
$\frac{h}{d+h-k}r! \prod_{i=1}^h p_i$ symbols of $F_{[h]}$ from each of the helper nodes, which achieves the cut-set bound \eqref{eq:cs1} with equality.
\end{enumerate}
\vspace*{.1in}

For every $i\in[h]$, define three sets $W_i^{(1)},W_i^{(2)}$ and $W_i$ as follows:
\begin{equation}\label{eq:dwi}
\begin{aligned}
W_i^{(1)} & := \Big\{\beta^{u_i t_i} \alpha_i^{u_i + q s_i}:
u_i=0,1,\dots,s_i-1; q=0,1,\dots,\frac{p_i-1}{s_i}-1 \Big\}, \\
W_i^{(2)} & := \Big\{\sum_{u_i=0}^{s_i-1} \beta^{u_i t_i} \alpha_i^{p_i-1} \Big\}, \\
W_i & := W_i^{(1)} \cup W_i^{(2)}.
\end{aligned}
\end{equation}
We will also use the following notation. Let
  \begin{gather*}
  \mathbi{u}_{\sim i} := (u_1,u_2,\dots,u_{i-1},u_{i+1},\dots,u_{h+1})\\ 
\mathbi{q}_{\sim i} := (q_1,q_2,\dots,q_{i-1},q_{i+1}, \dots,q_h).
  \end{gather*}
For every $i=1,2,\dots,h$, let
\begin{align*}
U_{\sim i} & := \{\mathbi{u}_{\sim i} : u_j = 0,1,\dots,s_j-1 \text{~for all~}
j\in \{1,2,\dots,h+1\} \backslash\{i\}\}, \\
Q_{\sim i} & := \{\mathbi{q}_{\sim i} : q_j = 0,1,\dots,p_j-1 \text{~for all~}
j\in [h] \backslash\{i\}\}.
\end{align*}

Finally, define the set
$S_i, i=1,2,\dots,h$
\begin{equation}\label{eq:DSi}
\begin{aligned}
S_i  := \bigcup_{\mathbi{u}_{\sim i} \in U_{\sim i}}
\bigcup_{\mathbi{q}_{\sim i} \in Q_{\sim i}}
W_i \beta^{(\sum_{j=1;j\ne i}^{h+1} u_j t_j)} \prod_{j\in [h]\backslash \{i\}} \alpha_j^{q_j},
\end{aligned}
\end{equation}
 which we will use to characterize the symbols downloaded for repairing the $i$-th node.
%(recall our notation \eqref{eq:dod}).
Again let $\cC^\bot=\text{\rm GRS}_{\mathbb{K}}(n,n-k,\Omega,v)$ be the dual code of $\cC$ \eqref{eq:grs}, where the coefficients $v=(v_1,\dots,v_n) \in (\mathbb{K}^*)^n$ are nonzero.
The theorem will follow from the following two lemmas.

\begin{lemma} \label{lem:Sh}
Node $c_1$ can be repaired from the set of symbols
$\{\trace_{\mathbb{K}/F_{[h]}}(\gamma v_jc_j):\gamma\in S_1,j\in \cR\}$.
Node $c_i,i=2,3,\dots,h$ can be repaired from the values $c_1,c_2,\dots,c_{i-1}$ together with the set of symbols 
$\{\trace_{\mathbb{K}/F_{[h]}}(\gamma v_jc_j):\gamma\in S_i,j\in \cR\}$.
\end{lemma}

\begin{lemma} \label{lem:iSH}
   \begin{equation}\label{eq:iSH}
\dim_{F_{[h]}}\big(\spun_{F_{[h]}}(S_1) + \spun_{F_{[h]}}(S_2)
        + \ldots + \spun_{F_{[h]}}(S_h)\big) = \frac{h}{d+h-k}r! \prod_{i=1}^h p_i.
   \end{equation}
\end{lemma}

Once these lemmas are established, the proof of the theorem can be completed as follows. According to
Lemma~\ref{lem:Sh}, to recover the values of the nodes $c_1,c_2,\dots,c_h$ it suffices to know the elements in the set
$D_j=\{\trace_{\mathbb{K}/F_{[h]}}(\gamma v_jc_j):\gamma\in \cup_{i=1}^h S_i\}$ from each of the helper nodes $\{c_j:j\in\cR\}$. 
To calculate the values of elements in the set $D_j$, it suffices to download the elements in the set
$\{\trace_{\mathbb{K}/F_{[h]}}(\gamma v_jc_j):\gamma\in B\}$, where the elements in $B$ form a basis of 
$\spun_{F_{[h]}}(S_1) + \spun_{F_{[h]}}(S_2)
+ \ldots + \spun_{F_{[h]}}(S_h)$ over $F_{[h]}$.
By Lemma~\ref{lem:iSH}, the count of these elements equals $\frac{h}{d+h-k}r! \prod_{i=1}^h p_i.$
Combining this with \eqref{eq:dim}, we conclude that the repair of $c_1,c_2,\dots,c_h$ from the helper nodes $\{c_j:j\in\cR\}$ indeed achieves the cut-set bound \eqref{eq:cs1}.

Moreover, it is clear from the proof that the repair field of the $h$-tuple $\{i_1,i_2,\dots,i_h\}$ is $\mathbb{F}_p(\{\alpha_j:j\in [n]\setminus \{i_1,i_2,\dots,i_h\}\})$. Therefore the largest common repair field for all the possible $h$-tuples of coordinates is ${\mathbb F}_p.$ This justifies the claim about the sub-packetization of our construction
made in \eqref{eq:sp}.
\end{IEEEproof}

Next let us prove Lemmas~\ref{lem:Sh} and \ref{lem:iSH}.

\vspace*{.1in}{\em Proof of Lemma~\ref{lem:Sh}:}
For every $i\in[h]$, define a field
\begin{equation}\label{eq:f[i]}
F_{[i]}:=\mathbb{F}_p(\{\alpha_j:j\in[n]\backslash[i]\}).
\end{equation}
Fix $i\in[h]$ and let us prove the lemma for the repair of the $i$-th node.
Let $h_i(x)$ be the annihilator polynomial of the set $\{\alpha_j: j\in[n]\setminus (\cR\cup[i]) \}$, i.e., 
\begin{equation}\label{eq:hi}
h_i(x)=\prod_{j\in[n]\setminus (\cR\cup[i])}(x-\alpha_j).
\end{equation}
Clearly, $\deg(x^t h_i(x))\leq s_i-1+n-(d+i)<n-k$ for all $t=0,1,\dots,s_i-1,$ so for any such $t$ we have
\begin{equation}\label{eq:mdi}
(v_1\alpha_1^t h_i(\alpha_1),\dots,v_n \alpha_n^t h_i(\alpha_n))\in\cC^\bot.
\end{equation}
These $s_i$ dual codewords will  be used to recover the $i$-th coordinate. 
Further, define a set $T_i$ whose elements will also be used to recover the $i$th coordinate:
\begin{equation}\label{eq:dTi}
T_i   :=  
\bigcup_{\mathbi{u}_{\sim i} \in U_{\sim i}}
\bigcup_{q_1 = 0}^{p_1-1}
\bigcup_{q_2 = 0}^{p_2-1}
\ldots
\bigcup_{q_{i-1} = 0}^{p_{i-1}-1}
\Big(W_i \beta^{(\sum_{j=1;j\ne i}^{h+1} u_j t_j)} \prod_{1\le j<i} \alpha_j^{q_j} \Big).
\end{equation}
It is easy to verify the following relation:
\begin{equation}\label{eq:sti}
S_i   = \bigcup_{q_{i+1} = 0}^{p_{i+1}-1}
\bigcup_{q_{i+2} = 0}^{p_{i+2}-1}
\dots
\bigcup_{q_h = 0}^{p_h-1}
T_i  \prod_{i< j \le h} \alpha_j^{q_j}.
\end{equation}

Let $c=(c_1,\dots,c_n)\in \cC$ be a codeword, and let us construct a repair scheme for the coordinate (node) $c_i$ using
the values $\{c_j:j\in \cR\cup\{1,2,\dots,i-1\}\}$. Rewrite \eqref{eq:mdi} as follows:
$$
\sum_{j=1}^n  v_j\alpha_j^t h_i(\alpha_j) c_j =0 \text{~for all~}  t=0,1,\dots,s_i-1.
$$
Computing the trace, we obtain
\begin{equation}\label{eq:its}
\sum_{j=1}^n \trace_{\mathbb{K}/F_{[i]}} (\gamma v_j\alpha_j^t h_i(\alpha_j) c_j) =0 \text{~for all~}  t=0,\dots,s_i-1 \text{~and all~} \gamma \in T_i.
\end{equation}
Let us write \eqref{eq:its} in the following form:
\begin{equation}\label{eq:tig}
\begin{aligned}
\trace_{\mathbb{K}/F_{[i]}} (\gamma \alpha_i^t v_i h_i(\alpha_i)c_i) & 
= - \sum_{j\neq i} \trace_{\mathbb{K}/F_{[i]}} (\gamma v_j \alpha_j^t h_i(\alpha_j)c_j) \\
& = - \sum_{j=1}^{i-1} \trace_{\mathbb{K}/F_{[i]}} (\gamma v_j \alpha_j^t h_i(\alpha_j)c_j)
- \sum_{j\in \cR} \trace_{\mathbb{K}/F_{[i]}} (\gamma v_j \alpha_j^t h_i(\alpha_j)c_j) \\
& = - \sum_{j=1}^{i-1} \trace_{\mathbb{K}/F_{[i]}} (\gamma v_j \alpha_j^t h_i(\alpha_j)c_j)
- \sum_{j\in \cR} \alpha_j^t h_i(\alpha_j) \trace_{\mathbb{K}/F_{[i]}} (\gamma v_j  c_j) \\
& \hspace*{1.5in} \text{~for all~} t=0,\dots,s_i-1 \text{~and all~} \gamma \in T_i,
\end{aligned}
\end{equation}
where the second equality follows from \eqref{eq:hi} and the third follows from the fact that the trace mapping $\trace_{\mathbb{K}/F_{[i]}}$ is $F_{[i]}$-linear, and that $\alpha_j\in F_{[i]}$
and $h_i(\alpha_j) \in F_{[i]}$ for all $j \in \cR$.

\vspace*{.1in}According to Prop.~\ref{prop:mt} in Appendix~\ref{ap:mt}, the set $\{\gamma\alpha_i^t:  t=0,1,\dots,s_i-1 ;\, \gamma\in T_i\}$ forms a basis\footnote{Note that the size of this set is $s_i|T_i|=(\prod_{j=1}^i p_j)( \prod_{m=1}^{h+1}s_m)$ 
which equals the extension degree $[{\mathbb K}:F_{[i]}]$ because of our definition of $s_{h+1}$ in \eqref{eq:sh+1}.} of  $\mathbb{K}$ over  $F_{[i]}$
and so does the set  $\{\gamma\alpha_i^t v_i h_i(\alpha_i):  t=0,1,\dots,s_i-1 ;\,  \gamma\in T_i\}$ (recall again that $v_i h_i(\alpha_i)\neq 0$).
Hence the value of $c_i$ can be calculated from 
$\{\trace_{\mathbb{K}/F_{[i]}} (\gamma \alpha_i^t v_i h_i(\alpha_i)c_i)
:t=0,1,\dots,s_i-1 ;\,  \gamma\in T_i \}$.

Using \eqref{eq:tig}, we conclude that the value of $c_i$ can be calculated from
the values of $c_1,c_2,\dots,c_{i-1}$ and the values of elements in the set
$\{\trace_{\mathbb{K}/F_{[i]}} (\gamma v_j  c_j):\gamma\in T_i, j\in\cR\}$.
The proof will be complete once we show that these elements can be found
%$\{\trace_{\mathbb{K}/F_{[i]}} (\gamma v_j  c_j):\gamma\in T_i, j\in\cR\}$ can be calculated 
from the elements in the set $\{\trace_{\mathbb{K}/F_{[h]}}(\gamma v_jc_j):\gamma\in S_i,j\in \cR\}$. This is an immediate consequence of \eqref{eq:trans} and equation \eqref{eq:sti}.
Indeed, observe that $F_{[i]}=F_{[h]}(\alpha_{i+1},\alpha_{i+2},\dots,\alpha_h),$ and that 
$\{\prod_{i< m \le h} \alpha_m^{q_m}: q_m=0,1,\dots,p_m-1, \forall i<m\le h\}$
 forms a basis of $F_{[i]}$ over $F_{[h]}$. Therefore, for every $\gamma\in T_i$ and every $j\in\cR$, the value of $\trace_{\mathbb{K}/F_{[i]}} (\gamma v_j  c_j)$ can be calculated from
    $$
    \Big\{\trace_{F_{[i]}/F_{[h]}}\Big(\trace_{\mathbb{K}/F_{[i]}} (\gamma v_j  c_j) \prod_{i< m \le h} \alpha_m^{q_m}\Big): q_m=0,1,\dots,p_m-1, \forall i<m\le h\Big\}.
    $$
Involving transitivity of the trace \eqref{eq:trans}, we see that
     \begin{align*}
\trace_{F_{[i]}/F_{[h]}}\Big(\trace_{\mathbb{K}/F_{[i]}} (\gamma v_j  c_j) \prod_{i< m \le h} \alpha_m^{q_m}\Big)
&= \trace_{F_{[i]}/F_{[h]}}(\trace_{\mathbb{K}/F_{[i]}} (\gamma v_j  c_j \prod_{i< m \le h} \alpha_m^{q_m} ) )\\
&= \trace_{\mathbb{K}/F_{[h]}} (\gamma v_j  c_j \prod_{i< m \le h} \alpha_m^{q_m} ),
     \end{align*}
where the first equality follows from the fact that $\alpha_m\in F_{[i]}$ for all $m>i.$
Therefore, for every $\gamma\in T_i$ and every $j\in\cR$, the value of $\trace_{\mathbb{K}/F_{[i]}} (\gamma v_j  c_j)$ can be calculated from
$$
\Big\{\trace_{\mathbb{K}/F_{[h]}} \Big(\gamma v_j  c_j \prod_{i< m \le h} \alpha_m^{q_m} \Big): q_m=0,1,\dots,p_m-1, \forall i<m\le h\Big\}
\subseteq \Big\{\trace_{\mathbb{K}/F_{[h]}}(\gamma v_jc_j):\gamma\in S_i,j\in \cR\Big\},
$$
 where the inclusion follows from \eqref{eq:sti}.
This establishes the needed fact, namely, that the elements in the set
$\{\trace_{\mathbb{K}/F_{[i]}} (\gamma v_j  c_j):\gamma\in T_i, j\in\cR\}$
can be calculated from 
$\{\trace_{\mathbb{K}/F_{[h]}}(\gamma v_jc_j):\gamma\in S_i,j\in \cR\}$, and completes the proof of Lemma~\ref{lem:Sh}. \hfill $\blacksquare$

\vspace*{.1in}{\em Proof of Lemma~\ref{lem:iSH}:}
We will prove the following more detailed claim (which implies the lemma):

\begin{claim}\label{claim1}
 For every $i\in[h]$,
\begin{equation}\label{eq:ck}
\dim_{F_{[h]}}\Big(\spun_{F_{[h]}}(S_1) + \spun_{F_{[h]}}(S_2)
+ \dots + \spun_{F_{[h]}}(S_i)\Big) = \frac{i}{d+i-k}r! \prod_{j=1}^h p_j.
\end{equation}
Moreover,  for every $i\in[h]$, there exist sets $B_i$ and $G_i$ that satisfy the following three conditions:
   \begin{enumerate}
  \item[$(i)$] $B_i$ is
a basis of $\spun_{F_{[h]}}(S_1) + \spun_{F_{[h]}}(S_2)
+ \dots + \spun_{F_{[h]}}(S_i)$ over $F_{[h]}$.
 \item[$(ii)$]
\begin{equation}\label{eq:rform}
B_i=
\bigcup_{u_{i+1} = 0}^{s_{i+1}-1}
\bigcup_{u_{i+2} = 0}^{s_{i+2}-1}
\dots
\bigcup_{u_{h+1} = 0}^{s_{h+1}-1}
\bigcup_{q_{i+1} = 0}^{p_{i+1}-1}
\bigcup_{q_{i+2} = 0}^{p_{i+2}-1}
\dots
\bigcup_{q_h = 0}^{p_h-1}
\Big( G_i \beta^{\sum_{j=i+1}^{h+1} u_j t_j} \prod_{i< j \le h} \alpha_j^{q_j} \Big).
\end{equation}
\item[$(iii)$]
\begin{equation}\label{eq:rcd}
G_i \subseteq \spun_{F_{[h]}} \Big(
\Big\{\beta^{\sum_{j=1}^i u_j t_j} \prod_{j=1}^i \alpha_j^{q_j}:
u_j=0,1,\dots,s_j-1 \text{~and~}
q_j=0,1,\dots,p_j-1
\text{~for all~} j\in[i]\Big\} \Big).
\end{equation}
\end{enumerate}
\end{claim}

{\em Proof of Claim~\ref{claim1}:}
Note that by \eqref{eq:tp} and \eqref{eq:rform},
\begin{equation}\label{eq:div}
|B_i|= \frac{r!}{t_{i+1}} \prod_{j=i+1}^h p_j |G_i|
\text{~for all~} i\in[h].
\end{equation}

We prove Claim \ref{claim1} by induction on $i$.
For $i=1$, we set $G_1=W_1$ and $B_1=S_1$, then conditions $(i)$--$(iii)$ are clearly satisfied. Moreover, it is easy to see that 
$|S_1|=\frac{1}{d+1-k}r! \prod_{j=1}^h p_j$. Together this establishes the induction base.

\vspace*{.1in}
Now let us prove the induction step. Fix $i>1$ and
 assume that the claim holds for $i-1$. By the induction hypothesis, \eqref{eq:ck} holds true, and there are a basis $B_{i-1}$ of $\spun_{F_{[h]}}(S_1) + \spun_{F_{[h]}}(S_2)
+ \dots + \spun_{F_{[h]}}(S_{i-1})$ over $F_{[h]}$ and a corresponding set $G_{i-1}$ that satisfy \eqref{eq:rform}-\eqref{eq:rcd}.
We have 
   $$
   |B_{i-1}| = \frac{i-1}{d+i-1-k}r! \prod_{j=1}^h p_j,
   $$
and so by \eqref{eq:div}
   $$
   |G_{i-1}| =\frac{i-1}{d+i-1-k} t_i \prod_{j=1}^{i-1} p_j 
=\frac{i-1}{d+i-1-k} \prod_{j=1}^{i-1} (s_j p_j).
   $$
Define the sets
\begin{align}
G_{[i]} & := \bigcup_{u_i=0}^{s_i-1} \bigcup_{q_i=0}^{p_i-1} G_{i-1} \beta^{u_i t_i} \alpha_i^{q_i},
\label{eq:dgi} \\
W_{[i]} & := \bigcup_{u_1=0}^{s_1-1}
\dots \bigcup_{u_{i-1}=0}^{s_{i-1}-1} \bigcup_{q_1=0}^{p_1-1}
\dots \bigcup_{q_{i-1}=0}^{p_{i-1}-1} \Big(W_i \beta^{\sum_{j=1}^{i-1} u_j t_j} \prod_{j=1}^{i-1} \alpha_j^{q_j}\Big). \label{eq:ww}
\end{align}
Let $G_i$ be a basis of 
$$
\spun_{F_{[h]}} (G_{[i]}) + \spun_{F_{[h]}} (W_{[i]})
$$
over $F_{[h]}$, and let $B_i$ be the set given by \eqref{eq:rform}.
It is clear that $G_i$ satisfies the condition \eqref{eq:rcd}. 

Next we show that 
$B_i$ is a basis of $\spun_{F_{[h]}}(S_1) + \spun_{F_{[h]}}(S_2) + \dots + \spun_{F_{[h]}}(S_i)$ over $F_{[h]}$.
By the induction hypothesis,
\begin{align}\label{eq:FS}
\spun_{F_{[h]}}(S_1) + &\spun_{F_{[h]}}(S_2) + \dots + \spun_{F_{[h]}}(S_{i-1})
\subseteq \spun_{F_{[h]}}(B_{i-1}) .
 \end{align}
 Now using \eqref{eq:rform}, we obtain
 \begin{align}
\spun_{F_{[h]}}(B_{i-1})&=  \spun_{F_{[h]}} \Big( 
\bigcup_{u_i = 0}^{s_i-1}
\bigcup_{u_{i+1} = 0}^{s_{i+1}-1}
\dots
\bigcup_{u_{h+1} = 0}^{s_{h+1}-1}
\bigcup_{q_i = 0}^{p_i-1}
\bigcup_{q_{i+1} = 0}^{p_{i+1}-1}
\dots
\bigcup_{q_h = 0}^{p_h-1}
\Big( G_{i-1} \beta^{\sum_{j=i}^{h+1} u_j t_j} \prod_{i\le j \le h} \alpha_j^{q_j} \Big) \Big) \nonumber \\
= & \spun_{F_{[h]}} \Big( 
\bigcup_{u_{i+1} = 0}^{s_{i+1}-1}
\bigcup_{u_{i+2} = 0}^{s_{i+2}-1}
\dots
\bigcup_{u_{h+1} = 0}^{s_{h+1}-1}
\bigcup_{q_{i+1} = 0}^{p_{i+1}-1}
\bigcup_{q_{i+2} = 0}^{p_{i+2}-1}
\dots
\bigcup_{q_h = 0}^{p_h-1}
\Big( G_{[i]} \beta^{\sum_{j=i+1}^{h+1} u_j t_j} \prod_{i< j \le h} \alpha_j^{q_j} \Big)
 \Big) \nonumber \\
\subseteq & \spun_{F_{[h]}} \Big( 
\bigcup_{u_{i+1} = 0}^{s_{i+1}-1}
\bigcup_{u_{i+2} = 0}^{s_{i+2}-1}
\dots
\bigcup_{u_{h+1} = 0}^{s_{h+1}-1}
\bigcup_{q_{i+1} = 0}^{p_{i+1}-1}
\bigcup_{q_{i+2} = 0}^{p_{i+2}-1}
\dots
\bigcup_{q_h = 0}^{p_h-1}
\Big( G_i \beta^{\sum_{j=i+1}^{h+1} u_j t_j} \prod_{i< j \le h} \alpha_j^{q_j} \Big)
 \Big) \nonumber \\
= & \spun_{F_{[h]}} ( B_i), \label{eq:eg}
\end{align}
where the second equality follows from \eqref{eq:dgi}; the inclusion on the third line follows from the definition of $G_i,$ and the last equality again follows from \eqref{eq:rform}.
According to \eqref{eq:DSi},
\begin{align}
& \spun_{F_{[h]}}(S_i) = 
\spun_{F_{[h]}} \Big(
\bigcup_{\mathbi{u}_{\sim i} \in U_{\sim i}}
\bigcup_{\mathbi{q}_{\sim i} \in Q_{\sim i}}
W_i \beta^{(\sum_{j=1;j\ne i}^{h+1} u_j t_j)} \prod_{j\in [h]\backslash \{i\}} \alpha_j^{q_j} \Big)
\nonumber \\
= & \bigcup_{u_{i+1} = 0}^{s_{i+1}-1}
\bigcup_{u_{i+2} = 0}^{s_{i+2}-1}
\dots
\bigcup_{u_{h+1} = 0}^{s_{h+1}-1}
\bigcup_{q_{i+1} = 0}^{p_{i+1}-1}
\bigcup_{q_{i+2} = 0}^{p_{i+2}-1}
\dots
\bigcup_{q_h = 0}^{p_h-1}
\Big( W_{[i]} \beta^{\sum_{j=i+1}^{h+1} u_j t_j} \prod_{i< j \le h} \alpha_j^{q_j} \Big) \nonumber \\
\subseteq & \spun_{F_{[h]}} \Big( 
\bigcup_{u_{i+1} = 0}^{s_{i+1}-1}
\bigcup_{u_{i+2} = 0}^{s_{i+2}-1}
\dots
\bigcup_{u_{h+1} = 0}^{s_{h+1}-1}
\bigcup_{q_{i+1} = 0}^{p_{i+1}-1}
\bigcup_{q_{i+2} = 0}^{p_{i+2}-1}
\dots
\bigcup_{q_h = 0}^{p_h-1}
\Big( G_i \beta^{\sum_{j=i+1}^{h+1} u_j t_j} \prod_{i< j \le h} \alpha_j^{q_j} \Big)
 \Big) \nonumber \\
= & \spun_{F_{[h]}}( B_i), \label{eq:ew}
\end{align}
where the second equality follows from \eqref{eq:ww}, and the inclusion follows from the definition of $G_i$.
Combining \eqref{eq:FS}, \eqref{eq:eg}, and \eqref{eq:ew}, we obtain that
\begin{equation}\label{eq:span}
\spun_{F_{[h]}}(S_1) + \spun_{F_{[h]}}(S_2) + \dots + \spun_{F_{[h]}}(S_i)
\subseteq \spun_{F_{[h]}}(B_i).
\end{equation}
Therefore,
$$
   |B_i| \ge \dim_{F_{[h]}}(\spun_{F_{[h]}}(S_1) + \spun_{F_{[h]}}(S_2)
+ \dots + \spun_{F_{[h]}}(S_i)).
$$
By Lemma~\ref{lem:Sh}, the number of symbols of $F_{[h]}$ downloaded from 
each of the helper nodes in order to repair
the nodes $c_1,c_2,\dots,c_i$, equals
   $
   \dim_{F_{[h]}}(\spun_{F_{[h]}}(S_1) + \spun_{F_{[h]}}(S_2)
+ \dots + \spun_{F_{[h]}}(S_i)).
     $
The cut-set bound implies that
\begin{equation}\label{eq:ge}
|B_i| \ge \dim_{F_{[h]}}(\spun_{F_{[h]}}(S_1) + \spun_{F_{[h]}}(S_2)
+ \dots + \spun_{F_{[h]}}(S_i)) \ge \frac{i}{d+i-k}r! \prod_{j=1}^h p_j.
\end{equation}

The proof of the induction step will be complete once we
show that
\begin{equation}\label{eq:le}
|B_i|\le \frac{i}{d+i-k}r! \prod_{j=1}^h p_j.
\end{equation}
Indeed, \eqref{eq:span}--\eqref{eq:le} together imply \eqref{eq:ck} and the needed fact that $B_i$ is a basis of $\spun_{F_{[h]}}(S_1) + \spun_{F_{[h]}}(S_2) + \dots + \spun_{F_{[h]}}(S_i)$ over $F_{[h]}$. 

Next let us prove \eqref{eq:le}.
From \eqref{eq:div}, this inequality will follow if we prove that
\begin{equation}\label{eq:ut}
|G_i|\le \frac{i}{d+i-k} \prod_{j=1}^i (s_jp_j).
\end{equation}
By the induction hypothesis and \eqref{eq:div}, we have
$
|G_{i-1}| = \frac{i-1}{d+i-1-k} \prod_{j=1}^{i-1} s_j p_j.
$
Combining this with \eqref{eq:dgi}--\eqref{eq:ww}, we obtain that
\begin{align*}
\left| G_{[i]} \right| & =
|G_{i-1}|s_i p_i = \frac{i-1}{d+i-1-k} \prod_{j=1}^i s_j p_j, \\
\left| W_{[i]} \right| & = |W_i|\prod_{j=1}^{i-1} s_j p_j = p_i\prod_{j=1}^{i-1} s_j p_j
= \frac{1}{d+i-k}\prod_{j=1}^i s_j p_j.
\end{align*}
Therefore,
\begin{equation}\label{eq:jh}
\begin{aligned}
|G_i| & =  \left| G_{[i]} \right|
+ \left| W_{[i]} \right| 
- \dim_{F_{[h]}} ( \spun_{F_{[h]}} (G_{[i]}) \cap
\spun_{F_{[h]}} (W_{[i]}) ) \\
& = \Big( \frac{i-1}{d+i-1-k} + \frac{1}{d+i-k} \Big) \prod_{j=1}^i (s_j p_j)
- \dim_{F_{[h]}} \Big( \spun_{F_{[h]}} (G_{[i]}) \cap
\spun_{F_{[h]}} (W_{[i]}) \Big).
\end{aligned}
\end{equation}
Since
$$
W_i \subseteq \spun_{F_{[h]}}\Big(\bigcup_{u_i=0}^{s_i-1} \bigcup_{q_i=0}^{p_i-1} \{\beta^{u_i t_i} \alpha_i^{q_i} \} \Big),
$$
we have
\begin{equation}\label{eq:cg}
G_{i-1}\odot W_i \subseteq \spun_{F_{[h]}} (G_{[i]}),
\end{equation}
where $\odot$ is defined in \eqref{eq:dod}.
According to \eqref{eq:rcd},
$$
G_{i-1} \subseteq \spun_{F_{[h]}}\Big(
\bigcup_{u_1=0}^{s_1-1}
\dots \bigcup_{u_{i-1}=0}^{s_{i-1}-1} \bigcup_{q_1=0}^{p_1-1}
\dots \bigcup_{q_{i-1}=0}^{p_{i-1}-1}  \beta^{\sum_{j=1}^{i-1} u_j t_j} \prod_{j=1}^{i-1} \alpha_j^{q_j} \Big),
$$
and consequently
$$
G_{i-1}\odot W_i \subseteq \spun_{F_{[h]}} (W_{[i]}).
$$
Combining this with \eqref{eq:cg}, we conclude that
$$
G_{i-1}\odot W_i \subseteq \spun_{F_{[h]}} (G_{[i]}) \cap
\spun_{F_{[h]}} (W_{[i]}).
$$
By the induction hypothesis, the elements in $B_{i-1}$ are linearly independent over $F_{[h]}$, and so are the elements in $G_{i-1}$. Using this together with the fact that the elements
in the set 
  $$
  \Big\{\beta^{\sum_{j=1}^i u_j t_j} \prod_{j=1}^i \alpha_j^{q_j}:
u_j=0,1,\dots,s_j-1 \text{~and~}
q_j=0,1,\dots,p_j-1
\text{~for all~} j\in[i]\Big\}
  $$
are linearly independent over $F_{[h]}$,
it is easy to see that the elements in $G_{i-1}\odot W_i$ are also linearly independent over $F_{[h]}$.
Therefore,
\begin{align*}
 \dim_{F_{[h]}} \Big( \spun_{F_{[h]}} (G_{[i]}) &\cap
\spun_{F_{[h]}} (W_{[i]}) \Big) \\
\ge & |G_{i-1}\odot W_i|= |G_{i-1}|\cdot |W_i| \\
= & \Big(\frac{i-1}{d+i-1-k} \prod_{j=1}^{i-1} (s_jp_j) \Big) p_i \\
= & \frac{i-1}{(d+i-1-k)(d+i-k)} \prod_{j=1}^i (s_jp_j) \\
= & \Big( \frac{i-1}{d+i-1-k} - \frac{i-1}{d+i-k} \Big) \prod_{j=1}^i (s_jp_j).
\end{align*}
Using this in \eqref{eq:jh}, we obtain that
\begin{align*}
|G_i| & \le \Big( \frac{i-1}{d+i-1-k} + \frac{1}{d+i-k} \Big) \prod_{j=1}^i s_j p_j
- \Big( \frac{i-1}{d+i-1-k} - \frac{i-1}{d+i-k} \Big) \prod_{j=1}^i s_jp_j  \\
& =  \frac{i}{d+i-k}  \prod_{j=1}^i s_jp_j.
\end{align*}
This establishes \eqref{eq:ut} and completes the proof of the claim. \hfill $\blacksquare$

%\begin{proposition}
%The sub-packetization of our construction is 
%$l=[\mathbb{K}:\mathbb{F}_p]=r!\prod_{i=1}^n p_i$, where $p_i$'s are the smallest $n$ distinct primes satisfying \eqref{eq:pmr}. 
%\end{proposition}
%To estimate the asymptotics of $l$ for $n\to\infty,$ 
%recall that our discussion of Dirichlet's prime number theorem in Section~\ref{Sect:rty} above implies that, for fixed $r$, $l= e^{(1+o(1)) n\log n}$. 

\appendices
\section{}\label{ap:PropT}
\begin{proposition}\label{prop:sumt}
For the set $T_{i_1}$ defined in \eqref{eq:defT1}, we have
$$
  \spun_{F_{i_1}}(T_{i_1}) + \spun_{F_{i_1}} (T_{i_1} \alpha_{i_1}) +\dots + 
\spun_{F_{i_1}} (T_{i_1} \alpha_{i_1}^{s_1-1} )=\mathbb{K},
$$
where $S \alpha := \{\gamma \alpha: \gamma \in S\}$,
and the operation $+$ is the Minkowski sum of sets, $T_1 + T_2 := \{\gamma_1+\gamma_2:\gamma_1\in T_1, \gamma_2\in T_2 \}.$ 
\end{proposition}
\begin{IEEEproof}
To establish the proposition, we will prove the following claim:
\begin{equation}\label{eq:cW}
  \spun_{F_{i_1}}(W_{i_1}) + \spun_{F_{i_1}} (W_{i_1} \alpha_{i_1}) +\dots + 
\spun_{F_{i_1}} (W_{i_1} \alpha_{i_1}^{s_1-1} )=\oplus_{u_1=0}^{s_1-1}\beta^{u_1}\mathbb{F}.
\end{equation}
Note that \eqref{eq:defT1} and \eqref{eq:cW} together imply that
\begin{align*}
  \spun_{F_{i_1}}(T_{i_1}) &+ \spun_{F_{i_1}} (T_{i_1} \alpha_{i_1}) +\dots + 
\spun_{F_{i_1}} (T_{i_1} \alpha_{i_1}^{s_1-1} ) \\
&=  \oplus_{u_1=0}^{s_1-1}\oplus_{u_2=0}^{s_2-1}\beta^{u_1+u_2s_1}\mathbb{F}\\
&= \oplus_{u=0}^{s-1} \beta^u \mathbb{F}\\
&= \mathbb{K},
\end{align*}
where the last equality follows from the fact that, on account of \eqref{eq:bbk}, the set $1,\beta,\dots,\beta^{s-1}$ forms a basis of $\mathbb{K}$ over $\mathbb{F}.$ Therefore the proposition indeed follows from \eqref{eq:cW}. 

Now we are left to prove \eqref{eq:cW}. This proof is close to the proof of Lemma 1 in \cite{Tamo17RS}, and we include it here for the completeness.

Let 
    $$
K:=\spun_{F_{i_1}}(W_{i_1}) + \spun_{F_{i_1}} (W_{i_1} \alpha_{i_1}) +\dots + 
\spun_{F_{i_1}} (W_{i_1} \alpha_{i_1}^{s_1-1} ).
   $$
Let us prove that $K=\oplus_{u_1=0}^{s_1-1}\beta^{u_1}\mathbb{F}.$ Clearly $K$ is a vector space over $F_{i_1}$, and by \eqref{eq:rFi1} we have $K\subseteq \oplus_{u_1=0}^{s_1-1}\beta^{u_1}\mathbb{F}$. 
Let us show the reverse inclusion, namely that $\oplus_{u_1=0}^{s_1-1}\beta^{u_1}\mathbb{F}\subseteq K$. More specifically, we will show that 
$\beta^{u_1}\mathbb{F} \subseteq K$ for all $u_1=0,1,\dots,s_1-1.$ 

We use induction on $u_1.$
For the induction base, let $u_1=0$, and let us show that the field $\mathbb{F}$ defined in \eqref{eq:defF} is contained in $K$.
In this case, we have $\alpha_{i_1}^{qs_1} \in W_{i_1}^{(1)}$ for all $0\leq q<\frac{p_{i_1}-1}{s_1}$.
Therefore $\alpha_{i_1}^{qs_1+j} \in W_{i_1}^{(1)} \alpha_{i_1}^j$ for all $0\leq q<\frac{p_{i_1}-1}{s_1}$.
As a result, $\alpha_{i_1}^{qs_1+j} \in K$ for all $0\le q<\frac{p_{i_1}-1}{s_1}$ and all $0 \le j \le s_1-1$.
In other words, 
\begin{equation}\label{eq:u1}
\alpha_{i_1}^t \in K \text{~for all~} t=0,1,\dots, p_{i_1}-2.
\end{equation}

Next we show that also $\alpha_{i_1}^{p_{i_1}-1} \in K$.
For every $t=1,\dots,s_1-1$ we have $0\le \lfloor \frac{p_{i_1}-1-t}{s_1} \rfloor <\frac{p_{i_1}-1}{s_1}$.
As a result,
$$
\beta^{t} \alpha_{i_1}^{t+ \lfloor \frac{p_{i_1}-1-t}{s_1} \rfloor s_1} \in W_{i_1}^{(1)}, \;
t=1,\dots,s_1-1.
$$
We obtain that, for each $t=1,\dots,s_1-1,$
$$
\beta^t \alpha_{i_1}^{p_{i_1}-1} = 
\beta^t \alpha_{i_1}^{t+ \lfloor \frac{p_{i_1}-1-t}{s_1} \rfloor s_1} 
\alpha_{i_1}^{p_{i_1}-1-t - \lfloor \frac{p_{i_1}-1-t}{s_1} \rfloor s_1}
 \in W_{i_1}^{(1)}  \alpha_{i_1}^{p_{i_1}-1-t - \lfloor \frac{p_{i_1}-1-t}{s_1} \rfloor s_1} \subseteq K.
$$
At the same time,
$$
\sum_{t=0}^{s_1 - 1}\beta^t \alpha_{i_1}^{p_{i_1}-1} \in W_{i_1}^{(2)} \subseteq K.
$$
 The last two statements together imply that
$$
\alpha_{i_1}^{p_{i_1}-1} = \sum_{t=0}^{s_1 - 1}
\beta^t \alpha_{i_1}^{p_{i_1}-1}
- \sum_{t=1}^{s_1 - 1}
\beta^t \alpha_{i_1}^{p_{i_1}-1}  \in K.
$$
Combining this with \eqref{eq:u1}, we conclude that
$\alpha_{i_1}^t \in K$ for all $t=0,1,\dots, p_{i_1}-1$.
Recall that $1,\alpha_{i_1},\dots,\alpha_{i_1}^{p_{i_1}-1}$  is a basis of $\mathbb{F}$ over $F_{i_1}$, and that $K$ is a vector space over $F_{i_1}$, so 
$\mathbb{F} \subseteq K$.
This establishes the induction base.

Now let us fix $u_1\ge 1$ and let us assume that $\beta^{u_1'}\mathbb{F} \subseteq K$ for all $u_1'<u_1.$ To prove the induction step, we need to show that $\beta^{u_1}\mathbb{F}\subseteq K$.
Mimicking the argument that led to \eqref{eq:u1}, we can easily show that
\begin{equation}\label{eq:ubeta}
\beta^{u_1} \alpha_{i_1}^{u_1+t} \in K
\text{~for all~} t=0,1,\dots, p_{i_1}-2.
\end{equation}
Let us show that \eqref{eq:ubeta} is also true for $t=p_{i_1}-1,$ i.e., that $\beta^{u_1} \alpha_{i_1}^{u_1+p_{i_1}-1} \in K$.
For every $1 \le t \le s_1-1-u_1$, we have $0\le \lfloor \frac{p_{i_1}-1-t}{s_1} \rfloor <\frac{p_{i_1}-1}{s_1}$.
As a result,
$$
\beta^{u_1+t} \alpha_{i_1}^{u_1+t+ \lfloor \frac{p_{i_1}-1-t}{s_1} \rfloor s_1} \in W_{i_1}^{(1)},\; t=1,\dots, s_1-1-u_1.
$$
Therefore, for all such $t$
\begin{equation}\label{eq:c1}
\beta^{u_1+t} \alpha_{i_1}^{u_1+p_{i_1}-1} = 
\beta^{u_1+t} \alpha_{i_1}^{u_1+t+ \lfloor \frac{p_{i_1}-1-t}{s_1} \rfloor s_1}
\alpha_{i_1}^{p_{i_1}-1-t - \lfloor \frac{p_{i_1}-1-t}{s_1} \rfloor s_1} 
  \in W_{i_1}  \alpha_{i_1}^{p_{i_1}-1-t - \lfloor \frac{p_{i_1}-1-t}{s_1} \rfloor s_1} \subseteq K
\end{equation}
By the induction hypothesis, $\beta^{u_1'} \mathbb{F} \subseteq K$ for all $u_1'=0,1,\dots,u_1-1$. As a result,
\begin{equation}\label{eq:c2}
\beta^{u_1'} \alpha_{i_1}^{u_1+p_{i_1}-1} \in K,\; u_1'=0,1,\dots,u_1-1.
\end{equation}
At the same time,
\begin{equation}\label{eq:c3}
\sum_{t=0}^{s_1 - 1}\beta^t \alpha_{i_1}^{u_1+p_{i_1}-1}
= \Big( \sum_{t=0}^{s_1 - 1}\beta^t \alpha_{i_1}^{p_{i_1}-1} \Big) \alpha_{i_1}^{u_1}
\in W_{i_1}^{(2)} \alpha_{i_1}^{u_1} \subseteq K.
\end{equation}
Combining \eqref{eq:c1}, \eqref{eq:c2} and \eqref{eq:c3}, we obtain that
$$
 \beta^{u_1} \alpha_{i_1}^{u_1+p_{i_1}-1} 
=  \sum_{t=0}^{s_1 - 1}\beta^t \alpha_{i_1}^{u_1+p_{i_1}-1}
- \sum_{u_1'=0}^{u_1-1} \beta^{u_1'} \alpha_{i_1}^{u_1+p_{i_1}-1} 
- \sum_{t=1}^{s_1-1-u_1} \beta^{u_1+t} \alpha_{i_1}^{u_1+p_{i_1}-1}
\in K.
$$
Now on account of \eqref{eq:ubeta} we can conclude that
$\beta^{u_1} \alpha_{i_1}^{u_1+t} \in K$  for all $ t =0,1,\dots, p_{i_1}-1$.
Therefore, $\beta^{u_1}\mathbb{F}\subseteq K$.
This establishes the induction step and completes the proof of the proposition.
\end{IEEEproof}

\section{}\label{ap:PropS}
\begin{proposition}\label{prop:sumS}
For the set $S_{i_2}$ defined in \eqref{eq:defSi}, we have
$$
  \spun_{F}(S_{i_2}) + \spun_{F} (S_{i_2} \alpha_{i_2}) +\dots + 
\spun_{F} (S_{i_2} \alpha_{i_2}^{s_2-1} )=\mathbb{K}.
$$
\end{proposition}
\begin{IEEEproof}
To establish the proposition, it suffices to prove that
\begin{equation}\label{eq:wrf}
  \spun_{F}(W_{i_2}) + \spun_{F} (W_{i_2} \alpha_{i_2}) +\dots + 
\spun_{F} (W_{i_2} \alpha_{i_2}^{s_2-1} )=
\oplus_{u_2=0}^{s_2-1}\beta^{u_2 s_1} F_{i_1},
\end{equation}
where $F_{i_1}$ is defined in \eqref{eq:fi1}.
Indeed, \eqref{eq:defSi} and \eqref{eq:wrf} together imply that
\begin{align*}
 \spun_{F}(S_{i_2}) + \spun_{F} (S_{i_2} \alpha_{i_2}) +\dots + 
\spun_{F} (S_{i_2} \alpha_{i_2}^{s_2-1} ) 
 &= \oplus_{u_1=0}^{s_1-1}\oplus_{u_2=0}^{s_2-1} \oplus_{q_1=0}^{p_{i_1}-1}\beta^{u_1 + u_2 s_1} \alpha_{i_1}^{q_1} F_{i_1} \\
&= \oplus_{u=0}^{s-1} \oplus_{q_1=0}^{p_{i_1}-1}\beta^u \alpha_{i_1}^{q_1} F_{i_1}\\
& = \oplus_{u=0}^{s-1} \beta^u \mathbb{F}\\
&= \mathbb{K},
\end{align*}
where the third equality follows from the fact that the set $1,\alpha_{i_1},\dots,\alpha_{i_1}^{p_{i_1}-1}$ forms a basis of $\mathbb{F}$ over $F_{i_1}$, and the last equality follows from the fact that the set $1,\beta,\dots,\beta^{s-1}$ forms a basis of $\mathbb{K}$ over $\mathbb{F}$ (see \eqref{eq:bbk}).
Thus the proposition indeed follows from \eqref{eq:wrf}.

The proof of \eqref{eq:wrf} is exactly the same as the proof of \eqref{eq:cW} (also the same as the proof of Lemma 1 in \cite{Tamo17RS}), and therefore we do not repeat it.
\end{IEEEproof}

\section{}\label{ap:mt}
\begin{proposition}\label{prop:mt}
For the set $T_i$ defined in \eqref{eq:dTi}, we have
$$
  \spun_{F_{[i]}}(T_i) + \spun_{F_{[i]}} (T_i \alpha_i) +\dots + 
\spun_{F_{[i]}} (T_i \alpha_i^{s_i-1} )=\mathbb{K}.
$$
\end{proposition}
\begin{IEEEproof}
To establish the proposition, it suffices to prove that
\begin{equation}\label{eq:wts}
  \spun_{F_{[i]}}(W_i) + \spun_{F_{[i]}} (W_i \alpha_i) +\dots + 
\spun_{F_{[i]}} (W_i \alpha_i^{s_i-1} )=
\oplus_{u_i=0}^{s_i-1}\beta^{u_i t_i} F_{[i-1]},
\end{equation}
where $W_i$ is defined in \eqref{eq:dwi}, and $F_{[i-1]}$ is defined in \eqref{eq:f[i]}.
Indeed, \eqref{eq:dTi} and \eqref{eq:wts} together imply that
\begin{align*}
   \spun_{F_{[i]}}(T_i) &+ \spun_{F_{[i]}} (T_i \alpha_i) +\dots + 
\spun_{F_{[i]}} (T_i \alpha_i^{s_i-1} ) \\
 = & \oplus_{\mathbi{u}_{\sim i} \in U_{\sim i}} 
\oplus_{q_1=0}^{p_1-1}
\oplus_{q_2=0}^{p_2-1}
\dots \oplus_{q_{i-1}=0}^{p_{i-1}-1}
\Big( \beta^{\sum_{j=1}^{i-1} u_j t_j+ \sum_{j=i+1}^{h+1} u_j t_j} \prod_{1\le j<i} \alpha_j^{q_j}
\big( \oplus_{u_i=0}^{s_i-1}\beta^{u_i t_i} F_{[i-1]} \big) \Big) \\
= &  
\oplus_{u_1=0}^{s_1-1}
\oplus_{u_2=0}^{s_2-1}
\dots \oplus_{u_{h+1}=0}^{s_{h+1}-1}
\oplus_{q_1=0}^{p_1-1}
\oplus_{q_2=0}^{p_2-1}
\dots \oplus_{q_{i-1}=0}^{p_{i-1}-1}
\Big( \beta^{\sum_{j=1}^{h+1} u_j t_j} \prod_{1\le j<i} \alpha_j^{q_j} F_{[i-1]}  \Big) \\
= &  
\oplus_{u=0}^{r!-1}
\oplus_{q_1=0}^{p_1-1}
\oplus_{q_2=0}^{p_2-1}
\dots \oplus_{q_{i-1}=0}^{p_{i-1}-1}
\Big( \beta^u \prod_{1\le j<i} \alpha_j^{q_j} F_{[i-1]}  \Big) \\
 =  & \oplus_{u=0}^{r!-1} \beta^u \mathbb{F} \\
= & \mathbb{K},
\end{align*}
where the third equality follows from \eqref{eq:nsu}; the fourth equality follows from the fact that for $j=2,3,\dots,h$, the set $1,\alpha_j,\dots,\alpha_j^{p_j-1}$ forms a basis of $F_{[j-1]}$ over $F_{[j]}$ and the fact that the set $1,\alpha_1,\dots,\alpha_1^{p_1-1}$ forms a basis of $\mathbb{F}$ over $F_{[1]}$, and the last equality follows from \eqref{eq:rst}.
%the fact that the set $1,\beta,\dots,\beta^{r!-1}$ forms a basis of $\mathbb{K}$ over $\mathbb{F}$ (see \eqref{eq:rst}).
Thus the proposition indeed follows from \eqref{eq:wts}.

The proof of \eqref{eq:wts} is exactly the same as the proof of \eqref{eq:cW} (also the same as the proof of Lemma 1 in \cite{Tamo17RS}), and therefore we do not repeat it.
\end{IEEEproof}

\bibliographystyle{IEEEtran}
\bibliography{repair}

\end{document}